# Beam extraction studies at 900 GeV using a channeling crystal

R.A. Carrigan, Jr.,[1] D. Chen,[*1] G. Jackson,[1] N. Mokhov,[1] C.T. Murphy,[1] S. Baker,[2]
A. Bogacz,[†3] D. Cline,[3] S. Ramachandran,[‡3] J. Rhoades,[‡‡3] J. Rosenzweig,[3] A. Asseev,[4]
V. Biryukov,[4] A. Taratin,[5] J. A. Ellison,[6] A. Khanzadeev,[7] T. Prokofieva,[7]
V. Samsonov,[7] G. Solodov,[7] B. Newberger,[§8] E. Tsyganov,[9] H.-J. Shih,[‖10] W. Gabella,[11]
B. Cox,[12] V. Golovatyuk,[12] and A. McManus[**12]

[1]*Fermi National Accelerator Laboratory, Batavia, Illinois 60510*
[2]*Argonne National Laboratory, Argonne, Illinois 60439*
[3]*University of California at Los Angeles, Los Angeles, California 90024*
[4]*Institute for High Energy Physics, Serpukhov, Russia*
[5]*Joint Institute for Nuclear Research, Dubna, Russia*
[6]*The University of New Mexico, Albuquerque, New Mexico 87131*
[7]*Petersburg Nuclear Physics Institute, Gatchina, Russia*
[8]*The University of Texas, Austin, Texas 78712*
[9]*Southwestern Medical Center, University of Texas at Dallas, Dallas, Texas 75235*
[10]*Superconducting Super Collider, Dallas, Texas 75237*
[11]*Vanderbilt University, Nashville, Tennessee 37235*
[12]*The University of Virginia, Charlottesville, Virginia 22901*

*Present address: AT&T, Middletown, NJ 07748.
†Present address: Thomas Jefferson National Accelerator Facility, Newport News, VA 23606.
‡Present address: University of Houston, Houston, TX 77204.
‡‡Present address: Fermilab, P. O. Box 500, Batavia, IL 60510.
§Present address: Winstead Sechrest & Minick, P.C., Austin TX 78701.
‖Present address: Affiliated Computer Services, Inc., Dallas TX 75204.
¶Present address: University and INFN of Lecce, Lecce, Italy.
**Present address: AT&T, West Long Branch, NJ 07764.

Luminosity-driven channeling extraction has been observed for the first time in a 900 GeV study at the Fermilab Tevatron. This experiment, Fermilab E853, demonstrated that useful TeV level beams can be extracted from a superconducting accelerator during high luminosity collider operations without unduly affecting the background at the collider detectors. Multi-turn extraction was found to increase significantly the efficiency of the process. The beam extraction efficiency was about 25%. Studies of time dependent effects found that the turn-to-turn structure was governed mainly by accelerator beam dynamics. An investigation of a pre-scatterer using the accelerator flying wire system showed that a fiber could produce a significant extracted flux, consistent with expectations. Based on these results, it is feasible to construct a parasitic 5-10 MHz proton beam from the Tevatron collider.

## I. INTRODUCTION

In the early 1990's a fixed target experiment was proposed for the Superconducting Super Collider[1] to study heavy-flavor physics using a tiny fraction of the 20 TeV circulating beam extracted with bent crystal channeling. Conventional methods of beam extraction at such high energies pose problems with no obvious cost-effective solutions. For example, electrostatic septa cannot easily provide the larger momentum kicks required for future trans-TeV accelerators. Since the new extraction scheme proposed for the SSC was not a proven technique, E853 at Fermilab was designed to study the feasibility of this approach.

The goals of E853 were to extract one million

900 GeV/c protons/second with $10^{12}$ protons circulating in the Tevatron, to study the extraction efficiency, to show that the luminosity lifetime of the circulating beam was not adversely affected, and to investigate the backgrounds created at the two Tevatron collider experiments. Losses at these major collider experiments, CDF (one sixth of the ring upstream for protons - see Fig. 1) and D0 (one sixth of the ring downstream), had to be kept to a tolerable level. A central concern for E853 operation was that losses be minimized so that the superconducting Tevatron magnets were not quenched.

There are significant constraints on where an inexpensive, parasitic extraction test can be placed within the Tevatron lattice. Any modification of the Tevatron vacuum system is expensive and requires rigorous attention to maintaining a very high vacuum. The C0 long straight section was chosen for E853 because it contained an existing 900 GeV abort system with an associated extraction line and dump that was not to be used in the 1994-1996 running period except at 150 GeV during collider injection tests. In addition there was free space along the abort beam line for instrumentation and cabling to a portakamp complex above ground for electronics and data acquisition.

Reliable simulations were required to interpret the experiment and explore ideas for halo generation and future applications. A simulation needs to cover all of the accelerator physics issues involved as well as the physics processes associated with channeling and multiple scattering in the crystal. Simulations created for the experiment have been discussed elsewhere by Bogacz et al.[2] and Biryukov[3].

The following sections describe the E853 experiment, the experimental apparatus, channeling behavior, measurements of the extraction rate and efficiency, time dependent and beam halo effects, RF, fiber-driven, and luminosity-driven extraction, as well as potential applications. Earlier presentations of the E853 results have appeared in Murphy et al.[4], Ramachandran's thesis[5], and Carrigan et al.[6]

## II. EXPERIMENT

The apparatus for E853 was placed at C0 where there was an existing Tevatron abort line (see Fig. 1). The experimental layout at C0 is shown schematically in Fig. 2. Protons from the circulating beam were kicked or diffused outward to a bent crystal that deflected the particles upward. From there they traveled along a simple beam line through a quadrupole doublet that was part of the Tevatron lattice to two diagnostic counter stations. Measurements consisted of determining the count rate in the extracted beam line as a function of crystal alignment, strength of the beam halo excitation, and other parameters.

The bent crystal was installed 60 m upstream of the C0 straight section center at the location of one of the kicker magnets for the 900 GeV C0 abort system (B48). Because the proton beam was on the outside of the centerline of the ring at B48, the crystal was placed on the outside of the ring (antiprotons were on the inside). The geometry is shown in Fig. 3(a).

The existing abort lattice was designed to provide an extraction system that would be fast (one turn) and clean so that it does not pose a problem for the string of superconducting magnets in the Tevatron[7]. The abort magnet string replaced by the E853 crystal consisted of four 1.8 m long kicker modules with peak fields of 3.7 kG and a rise-time of 1.5 μsec giving a total vertical deflection of 640 μrad. Removal of the most upstream kicker module provided sufficient space for the crystal goniometer.

Extraction consists of two parts: a vertical kick into the field-free region of a string of Lambertson magnets, as shown in Fig. 3(b), and horizontal separation of the circulating beam from the extracted beam by the Lambertsons. In the Tevatron lattice missing half-dipoles at B48-3 and at C11 establish a three-bend magnetic dogleg and thereby provide an effective four mrad horizontal kick at B48 so that the abort line can clear the magnets at the downstream end of the straight section. The missing bend in the accelerator ring resulting from the removal of the half-dipole is supplied by the warm Lambertson magnets and C-magnets in the straight section itself. The field-free and field regions of the Lambertsons then provide the initial horizontal separation between the extracted and circulating beams. For E853 the 640 μrad crystal bend deflected the beam incident on the crystal up into the field-free region of the Lambertson magnets.

Since E853 was carried out parasitically to the collider program, it was not possible to optimize the accelerator lattice for channeling extraction. For example, it would have been desirable to have more dispersion at the crystal so that longitudinal noise excitation could have been carried out. The accelerator parameters are given in Table I for the two types of stores in which E853 had a study session: low-intensity proton-only stores of about four hours duration in which E853 was the only user, and high-intensity $\bar{p}p$ stores in which E853 took data parasitically during the last few hours. There were a total of 30 sessions in E853.

TABLE I. Accelerator pararameters for the two types of stores in which E853 took data. D is the horizontal dispersion, $\varepsilon$ is the one-$\sigma$ emittance, $\beta$ and $\alpha$ are the standard parameters describing the shape of the emittance ellipse, and $\nu_x$ and $\nu_y$ are the horizontal and vertical tunes of the accelerator. x-p and y-p are the horizontal and vertical positions of the protons at the crystal relative to the central orbit of the Tevatron.

| Parameter | Proton-only | $\bar{p}p$ | Units |
|---|---|---|---|
| $\varepsilon$ | 3.33 | 6.66 | $\pi$ mm-mrad |
| $\beta_x$ | 100. | 104. | m |
| $\beta_y$ | 29. | 23. | m |
| $\alpha_x$ | -0.04 | 0.13 | |
| $\alpha_y$ | 0.01 | 0.10 | |
| D | 1.926 | 1.887 | |
| $\sigma_x$ | 0.75 | 0.91 | mm |
| $\sigma_y$ | 0.33 | 0.41 | mm |
| $\sigma_{\Theta x}$ | 6.1 | 8.5 | $\mu$ rad |
| $\sigma_{\Theta y}$ | 11.3 | 18.0 | $\mu$ rad |
| $\nu_x$ | 20.585 | 20.585 | |
| $\nu_y$ | 20.574 | 20.574 | |
| x-p | 0 | 1.99 | mm |
| x-$\bar{p}$ | | -1.99 | mm |
| y-p | 0 | -0.75 | mm |
| y-$\bar{p}$ | | 0.75 | mm |

After the Lambertson magnets the extracted beam traversed two instrumented air gaps approximately 100 meters downstream of the crystal and then entered a beam dump. The air gaps, separated by 40 meters, were each instrumented with several scintillators. A fluorescent screen coupled to a CCD camera in the first air gap also provided a digital readout of the beam-profile for run-time diagnostics. There was already a segmented-wire ion chamber (SWIC) in the second air gap as well as a toroid to measure beam current.

In the collider mode, the Tevatron typically operated with six bunches of protons and six bunches of antiprotons circulating so that a proton bunch passed the crystal every 3.49 μsec. The proton and antiproton orbits were separated at most points in the ring by an electrostatic separator system. A typical bunch length was 2 ns. Characteristically the beam circulated in the Tevatron for 10 to 20 hours as the beam slowly decayed because of such factors as losses at the collision points, gas scattering, and scraping.

Under most circumstances the Main Ring Accelerator located just above the Tevatron was also operating to build an antiproton stack during collider operations. It cycled once every several seconds and produced large backgrounds in the E853 detectors, so it was necessary to gate off the electronics when protons were in the Main Ring. However magnetic effects of bus work and the magnet ramping affected Tevatron orbits throughout the Main Ring cycle.

Two techniques were used for extracting the beam in E853. In one approach a fast kicker magnet was used to induce a betatron oscillation in one of the six bunches. The edge of this bunch then spilled out on the crystal over the next 50 - 100 turns. This mode of operation is called "kick mode". The nominal value used for the kicker voltage was 10 KV, creating a maximum step size at the crystal of 0.5 mm.

The principal purpose of kick mode was to get beam incident deep in the crystal so that the beam was beyond any surface irregularities of the crystal edge (which can be of the order of a micron) and beyond the effective surface layer created by the horizontal angular alignment error, which was of the order of 0.3 mrad. Irregularities on the surface of the crystal and crystal plane misalignment had the same effect as that of electrostatic septa wires, scattering the beam as the beam encountered an irregular and possibly amorphous layer instead of uniform planes.

The disadvantage of kick mode was that the beam intensity and emittance degraded at a rate that made it unacceptable for parasitic extraction during collider operations. Kick mode was used only in proton-only stores.

Kick mode was employed in the initial sessions of E853. The basic technique was to move the crystal in to a predetermined horizontal location close to the beam (4.5 to 6 $\sigma_x$ from the beam center). The beam was then kicked horizontally a number of times by a fast kicker at E17 (see Fig. 1). At this distance from the crystal, the beam density was so small that no beam was observed to interact with the crystal on the first few kicks.

However, after several hundred turns following a kick, the beam had grown in size as non-linearities in the machine gradually spread the beam to fill much of the phase space mapped out by the betatron oscillation. After about six such kicks, the beam size had grown by a factor 1.7 in the horizontal plane and a factor 1.2 in the vertical plane (resulting from the horizontal/vertical coupling in the Tevatron). In the kicks which followed, an equilibrium state persisted in which the rising edge of the Gaussian beam distribution led to a nearly constant loss of about $2 \cdot 10^9$ protons/kick, and extraction by the crystal was observed.

The second extraction technique was called "diffusion mode", relying on either natural or stimulated diffusion during proton-only stores and $\bar{p}p$ collisions parasitic to collider operations. This diffusion was due to such effects as beam-gas scattering, beam-beam scattering and tune shift effects, magnetic field ripple, and imposed RF noise. In this mode, the crystal was slowly moved into the tail of the beam halo. As the crystal approached the beam, the extraction rate rose rapidly, as illustrated in Fig. 4. Movement of the crystal was halted when an extraction rate adequate for the studies had been achieved. When operating parasitic to collider operations, this mode did not seriously affect the circulating proton beam lifetime.

Protons were steadily resupplied onto the crystal either by natural diffusion, artificially applied RF noise, or $\bar{p}p$ collisions. Beam which did not channel on the first encounter with the crystal underwent multiple scattering and got another chance to enter the crystal on successive turns, if it did not interact in the crystal. The most effective instrumentation for beam measurement in this mode was the scintillation counter system set at voltage levels appropriate for single-particle identification.

A horizontal damper located at F11 was also used in two E853 sessions to introduce RF noise in transverse phase space and thereby stimulate diffusion and increase the extraction rate. This characteristically decreased the beam lifetime substantially although it was still long enough for a typical study session. Significantly higher extraction rates were observed with the noise on. This approach could not be used for parasitic extraction owing to its destructive effect on the circulating beam.

### III. APPARATUS

This section reviews the main components of E853. The operation of the Tevatron, which was at the heart of the experiment has already been described. The other principal components of the apparatus were the channeling crystal with its associated goniometer and the diagnostic instrumentation to carry out the experiment. Both special instrumentation and the existing Tevatron diagnostics were required to study the circulating and extracted beams. Finally, data collection was provided by an on-line computer system as well as the extensive accelerator ACNET control utility.

#### A. Bent crystal operation

Channeling requires a well-characterized crystal, a device to bend the crystal, and a precise and flexible goniometer. These systems are discussed in the following paragraphs.

##### 1. Crystal

The "dislocation-free" silicon crystal used for E853 was obtained from Wacker Corporation after crystals from several sources were evaluated for dislocation density using a two crystal x-ray diffraction spectrometer discussed in Baublis *et al.*[8] The crystal had a diffraction line width of (4.5±0.5) μrad which confirmed that it was dislocation-free. For the experiment several 39 mm long, 3 mm high, 9 mm wide crystals were prepared with different planar orientations.

The crystal used in the experiment was cut so that a (111) plane was aligned to within 1000 µrad of the crystal face. A more comprehensive view of the planar alignment for one of the crystals is shown in Fig. 6 of Baublis *et al*. The side facing the beam was polished to be optically flat within half a wavelength. The results of this process are illustrated in Fig. 7 of Baublis *et al*.

*2. Bender*

A 4-point bender was chosen for E853 to produce a uniform bend. A drawing of the bender with the crystal and alignment mirror is shown in Fig. 5. Note that the crystal had straight overhangs of 7.5 mm beyond the outer points of support for the bend. More details on the design aspect of the bender can be found elsewhere[9].

The crystal was pre-bent up through an angle of 642 µrad. The bend angle was set to within 5 µrad using two independent techniques: laser interferometry and auto collimation. A small mirror was carefully aligned with the top surface of the upstream end of the crystal prior to installation. The crystal was positioned in the Tevatron using survey monuments on the outside of the goniometer. That alignment was checked by auto-collimation using the upstream reference mirror. The upstream top face orientation of the crystal was estimated to be within 150 µrad of the Tevatron beam based on the effect of refraction in the viewing window, 10 µrad surveying errors, and other factors.

Both the horizontal and vertical angle initial alignments in the accelerator were assisted greatly by relating the upstream and downstream ends of the two crystal surfaces (only 40 mm long) to the survey monuments external to the 1-m long goniometer tube using the Coordinate Measuring Machine of the Fermilab Silicon Detector Facility. This device has an accuracy of 2 µm globally, leading to an uncertainty of about 100 µrad in the angles of the crystal with respect to the external monuments. The optical techniques for measuring the external fiducials during installation in the accelerator are accurate only to 100 µm, implying an additional uncertainty of 100 µrad over the 1 m length.

*3. Goniometer*

For channeling to occur, the vertical angular orientation of the channeling crystal must be aligned with the beam angle. Since the Tevatron beam has a vertical angular divergence of 11 µrad in proton-only stores and the critical angle of the crystal at 900 GeV is 5.2 µrad, the crystal alignment must be within of order 20 µrad for channeling to occur. A special goniometer was designed for E853 to position and align the crystal quickly inside the vacuum system of the Tevatron. Precision stepping motors outside the pipe at each end of the vacuum pipe allowed alignment of the crystal with four degrees of freedom (x, $\Theta_h$, y, $\Theta_v$).

The crystal was mounted at the upstream end of a 1 m long beam pipe with articulating bellows at the ends. Because of the upstream location of the crystal the angular motion was provided solely by the downstream motors and the crystal turned around its own center. The motors produced step sizes of 2.5 µm which translated to 2.5 µrad in angle with the 1 m lever arm of the goniometer. The accelerator control system was used to control the goniometer. A schematic of the goniometer is shown in Fig. 6(a).

A horizontal retractor controlled by a fifth motor was used to move the crystal holder in and out of the beam. The crystal was completely outside of the Tevatron aperture when not in use. The retractor also preserved the crystal orientation over many excursions of the crystal holder out of the beam pipe with a reproducibility of 15 µrad. The retraction arrangement is shown in Fig. 6(b).

Extensive reproducibility studies of the goniometer motions were carried out. These were done for different load conditions to simulate the real conditions with the meter-long insert attached to the motion tables. The minimum step size on the x, y motion stepping motors and on the motor used by the retraction mechanism were verified using a high resolution (1 µm) optical encoder. The backlash error was of the order of 3 to 4 µm.

**B. Extraction line instrumentation**

The instrumentation for beam detection and measurement was placed in two air gaps along the abort line. Eight plastic scintillation counters of various shapes and sizes were used to monitor the beam flux close to the crystal and along the extraction line. The operating voltages for single particle counting were determined with cosmic rays and beam coincidence studies. For diffusion mode the counters were operated as single particle counters. Count rates were limited to somewhat less than the bunch passage frequency (300 Khz for 6 bunches).

For kick mode a significant fraction of the circulating beam was extracted down the abort line in the first few turns so that it was not possible to count individual particles. Typically $2 \cdot 10^9$ protons were lost per kick. Assuming an extraction efficiency of 30% on the order of $10^9$ protons could come down the extraction line in a single RF bucket. This flux could saturate the counter itself or the analog to digital converter (ADC) used for charge integration. Kick mode voltages were set lower to avoid these problems. The scintillators acted as calorimeters, and their pulse heights were recorded.

Three principal counter combinations were used. Interaction counters below and slightly downstream of the crystal were employed to monitor interactions of the beam with the crystal. Two counters in the first air gap formed a coincidence called AG1. This was often placed in coincidence with a counter in the second air gap such as one referred to as CAL.

In kick mode ADC readings for the various counters were taken for the second revolution of the proton bunch around the ring after the beam was kicked (the beam was on the far side on the first turn after the kick). Typically several scintillator signals were also recorded for many turns with digital oscilloscopes.

Because the C0 abort line was routinely used for 150 GeV extraction of the entire beam during collider setup periods, the counters at the air gaps were mounted on horizontal linear motion stages in order to allow them to be removed remotely when not in use. A vertical finger counter (width = 1.6 mm) on the second air gap stage was used to study the horizontal beam width while a horizontal finger counter (height = 0.8 mm) was independently mounted on a vertical motion in the first air gap to measure the vertical profile. The ACNET control system was used to operate the stages.

The extracted beam was also monitored with a chromox fluorescent screen (from Morgan Matroc Co.) at the first air gap. A conventional segmented wire ion chamber (SWIC) with 1 mm wire spacing in x and y already existed in the second air gap. In kick mode both of these devices provided information on the beam size and position. The fluorescent screen and the SWIC were not useful in the diffusion mode because there was not a high enough flux of beam particles to produce signals. The fluorescent screen was viewed with a CCD camera connected to a video monitor and tape recorder. Video tapes were used to look at time dependent effects with scales greater than 15 ms in the kick mode.

The screen image of the beam was also captured and stored with the help of a "frame grabber". The images were analyzed later to determine the beam distribution in both the horizontal and vertical projections. Fig. 7 shows typical "frame grabber" views of the beam spot for the kick mode. Despite problems with saturation and calibration, the screen provided a useful visual tool to guide one during the initial setup for crystal alignment. The fluorescent screen was also useful in determining the extracted beam position with respect to the aperture of the Lambertson magnets.

**C. Accelerator instrumentation**

The characteristics of the circulating beam and the loss rates at the collider detectors were measured using existing Tevatron diagnostics. Electromagnetic beam position monitors (BPMs) in the accelerator ring continuously tracked the position of each bunch of the circulating beam at many points around the accelerator and were used to determine the distance of the proton beam from the crystal. Resistive wall monitors and DC current transducers tracked the total circulating intensity. These signals, averaged over several turns, were used to determine the total loss rate from the beam, not just that resulting from the crystal. They were also useful quantities for normalizing the extraction rate in the diffusion mode.

Beam profile measurements in the Tevatron were done with a system of three "flying wires" at E11 and E17. A flying wire is a 30 μm diameter carbon fiber that moves through the beam with a speed of ≈ 5 m/sec. The secondary particles from the interaction of the beam with the filament are detected as a function of time using a scintillator telescope. This gives the beam intensity profile directly.

Vertical and horizontal beam profiles were taken in kick mode with the flying wires before and after beam growth with the first six kicks. With the emittances determined at one location, it was straight-forward to calculate the beam divergences at the crystal knowing all the relevant lattice Twiss functions.

Loss rates at the collider detectors were measured using permanent counters located at B0 (CDF) and D0. For ordinary Tevatron collider operations the circulating beam was scraped until these count rates were below standard operating upper limits. Data from all these systems were available at E853 through the accelerator ACNET system.

## IV. CHANNELING BEHAVIOR

While E853 was the highest energy channeling measurement yet performed, it was undertaken to look at channeling extraction, not the fundamentals of the channeling process. Because the energy was high, the channeling critical angle was small, 5.2 µrad at 900 GeV. As a consequence, it was difficult to measure the size of this angle directly. Nevertheless, information from E853 confirms that the channeling process continues to behave consistently in the TeV range.

### A. Crystal angular alignment

#### 1. Vertical angle alignment

For successful extraction the channeling plane of the crystal must be aligned with the circulating accelerator beam. The angular width of the $\Theta_V$ scan (varying the vertical angle of the crystal) is determined by three factors. One is the channeling critical angle, another is the angular divergence of the accelerator beam, $\sigma_{\Theta_V}$, and the third is the impact of multiple passes through the crystal on the angular divergence of the beam halo.

At the crystal $\sigma_{\Theta_V}$ was typically 18 µrad during collider-mode stores. The rms multiple scattering for one pass through the crystal was 10.8 µrad. Characteristically in the diffusion mode a proton passed through the crystal several times before it was properly aligned for channeling. The effective multiple scattering angle was proportional to the square root of the number of passes through the crystal.

The lower panel of Fig. 8 shows a typical crystal $\Theta_V$ extraction curve in diffusion mode during a collider store. The simulation[3] predicts $\sigma_{\Theta_V}$ to be 21 to 24 µrad compared to 32 µrad measured in Fig. 8. Fitted values for various sessions ranged from 28 to 37 µrad. The discrepancy between the simulation and experiment arises because the simulation took the beam divergence to be 11.5 µrad, whereas the beam divergence in collider mode was 18 µrad (see Table I).

In kick mode, data was taken for the $\Theta_V$ scan and other scans only on the first pass of the kicked bunch through the crystal (the second turn after the kick - see Section VII below). Therefore, the $\Theta_V$ curve should be narrower than in diffusion mode because there was no multiple scattering. The upper panel of Fig. 8 shows a kick mode vertical scan gated on the second turn for which $\sigma_{\Theta_V}$ = 18 µrad, clearly illustrating the suppression of multi-pass multiple scattering. Because of the beam growth resulting from the kicks, this width was expected to be somewhat greater than 15 µrad.

#### 2. Horizontal angle alignment

Before the experiment began there was concern that the horizontal angle of the crystal might have to be aligned very accurately in the diffusion mode so that the edge was quite parallel to the beam. Fortunately, multi-turn extraction in the diffusion mode obviates this problem, since multiple scattering from the portion of the crystal projecting out into the beam scatters particles out to larger radii on later turns.

Beyond this concern there were several impacts of horizontal angular scans. As the crystal rotated in $\Theta_h$ starting from a well-aligned angle ($\Theta_h$ = 0), either the downstream or the upstream end moved closer to the beam. For example, a 10 mrad change in $\Theta_h$ moved both ends of the crystal by 200 μm in opposite directions. As a result of the end moving into the beam the interaction counter rate rose. This behavior is illustrated in Fig. 9. The time order of the data was from right to left. As $\Theta_h$ decreased from its starting value of 8 mrad, the interaction rate is observed to decrease because as the upstream end of the crystal moved out of the beam it was intercepting a region from which most of the particles had been removed at the previous angular setting, and diffusion had not yet had a long enough time to repopulate the void.

This effect was anticipated, for it had been observed that when the whole crystal was translated away from the beam by 200 μm, it took some minutes for diffusing beam to reestablish a constant interaction rate. When the goniometer reached $\Theta_h = 0$, the downstream end of the crystal began to move into the beam and intercept increasingly repopulated regions as the angle continued to change; therefore, the interaction rate rose steadily.

We take the interaction rate to be proportional to the amount of beam incident on the crystal, and the quantity of interest to plot as a function of $\Theta_h$ is the beam extracted divided by the background-subtracted (see Section VI below) interaction counter rate, as shown in Fig. 10. When the upstream end of the crystal was protruding into the beam (positive $\Theta_h$), some initially channeled portion of the beam should have escaped through the side facing the beam before it was fully deflected. When the downstream end pointed in (negative $\Theta_h$), the unaligned portion of the crystal (the downstream end, which was bent upwards) intercepted the beam first so that less beam should have been deflected. Both effects are observed in Fig. 10.

Biryukov[3] studied the horizontal alignment with his simulation program in some detail with the hope that the horizontal alignment curve could be used to probe questions of the diffusion rate and imperfections on the beam facing surface. For comparison the Biryukov simulation for the E853 crystal and bender is also shown on Fig. 10. The simulation took into account the real geometry of the crystal including straight sections and variable curvature plus an assumed 10 μm thick amorphous layer on the surface. The two distributions are in fairly good agreement.

For kick mode a uniform 20 μrad change in $\Theta_v$ was observed with the CCD camera over a 20 mrad change in $\Theta_h$ around $\Theta_h = 0$. This change was in line with the change predicted for crystal distortions resulting from the bender. As the downstream crystal end moved away from the beam to angles beyond 10 mrad there was a slight indication that the deflection was decreasing. For a 0.5 mm kick this effect should have become significant for $\Theta_h > 20$ mrad.

### B. Extracted beam size

The beam optics in the vertical plane of the extracted beam was point to parallel from the crystal to the air gap region so that the beam height was directly related to the critical angle. Because of the optics, the accelerator beam height contributed little to the extracted beam height. The vertical beam profile was measured with a 0.8 mm high finger counter in the first air gap. The smearing of the vertical beam height distribution because of the finite counter width was $0.8$ mm/$\sqrt{8.5}$. Fig. 11 shows the distribution for the most carefully measured finger scan. The vertical beam distribution gave a raw width of $\sigma_v = 0.32$ mm. Factoring out the finite counter width gives $\sigma_v = 0.25$ mm. The expected width based on the beam optics and the critical angle was $\sigma_v = 0.23$ mm.

Alternatively, one can calculate a value for the critical angle directly by noting that $\sigma_v \approx M_{12} \Psi_c$ where $\Psi_c$ is the critical angle, and $M_{12}$, the beam transfer matrix element from the crystal to air gap 1, was 41.3 m. Substituting the measured extracted beam height gives $\Psi_c = 6.0$ μrad, compared to the expected value of 5 μrad. The apparent good agreement may be fortuitous. While no error has been assigned, it would be dominated in part by uncertainty in the finger counter angular alignment, which would make the effective counter width greater than 0.8 mm. Rather, the point is that the vertical distribution was consistent with the magnitude of the critical angle.

Because of the optics, the horizontal beam distribution in the air gaps was dominated by the halo angular distribution that reached the crystal. The horizontal size of the source at the crystal was also quite small in the diffusion mode so that $\sigma_h \approx 132\ \sigma_{\Theta_h}$ where 132 m is the appropriate beam transfer matrix element.

Only three plausible horizontal distributions were measured because of time pressure and problems with the counter. A proton-only store gave $\sigma_h = 2.8\pm0.5$ mm while two colliding beam stores gave $\sigma_h = 1.1\pm0.2$ mm and $1.2\pm0.2$ mm. The vertical finger counter in air gap 2 was 1.6 mm wide. When the effect of the width is factored out, the horizontal width ranged from 1.0 to 2.8 mm which suggests values for the halo angular distribution on the crystal of 8 to 21 µrad. The lower value is consistent with the angular distribution based on multi-pass multiple scattering extraction.

### C. Dechanneling

A flat dechanneling tail is clearly visible in the y finger distribution in Fig. 11 and in the CCD images of Fig. 7. The tail is cut off on the left side of Fig. 11 by the steel of the Lambertson magnet. The percentage of the particles in the visible tail is 20% of the peak. The simulation[3] predicted $26\pm2\%$. Given the complexity of the dechanneling process for the accelerator beam halo, the agreement between the experiment and the simulation is adequate.

## V. EXTRACTION RATES

A principal purpose of this experiment was to study absolute extraction rates under various conditions. We have measured extraction rates under three conditions: extraction driven by natural diffusion during proton-only stores, RF noise-driven diffusion during a proton-only store, and luminosity-driven extraction during proton-antiproton stores.

Direct digital counting measurements of the beam extracted down the beam line required several corrections. The pair of counters in coincidence in the first air gap which counted the extracted beam was corrected for the 90% efficiency of the pair. Accidental coincidences were negligible because of the low rate of non-extraction backgrounds. Small corrections for multiple particles extracted from the same RF bucket which the scintillators would count as one particle were calculated by Poisson-statistics techniques and were empirically verified in three of the sessions by recording the pulse heights in each of the counters for 10,000 events. In the worst case, this correction raised the extraction rate by 70%, but more typically was a 10% correction.

The finger counter distributions were used to make other small corrections for ambient background and for the dechanneling tail, which was subtracted from the extraction rate because in a real application of this technique this tail would not survive the rest of the beam line. This correction typically reduces the extraction rate 15%.

In a typical proton-only store, $10^{11}$ protons were circulating in six bunches. In this mode, the maximum extraction rate achieved was 200 kHz. Higher rates could have been achieved by moving the crystal even closer to the beam, but with only six bunches, a rate of 287 kHz corresponded to extracting on average one proton per bunch, and the counters could not count more than one particle per bunch.

To mitigate this limitation, a special store was arranged with $10^{11}$ protons circulating in 84 bunches. Additional diffusion was induced by transverse RF horizontal noise using the electrical damper located at F11, creating an rms diffusion rate at the crystal of 0.04 µm per turn. The extraction rate was greater than 450 kHz.

In the luminosity-driven stores, typically $10^{12}$ protons were circulating in six bunches. The maximum extraction rate achieved was 150 kHz. In this mode the limitation was the impact of particles scattered from the crystal in creating backgrounds for the operating collider experiments. Although the CDF experiment received no measurable background from the crystal, the D0 "lost proton" monitor was sensitive to scattering from the crystal. D0 was usually already running at 80% of the conservative upper limit set by that experiment before the crystal was moved close to the beam and reached the limit when the extraction rate was between 50 and 150 kHz.

This limitation was removed during a special store with 36 proton bunches and 3 antiproton bunches during which D0 was not taking data. There were $3\cdot10^{12}$ protons circulating, and an extraction rate of 900 kHz was achieved. The D0 lost proton monitor exceeded its upper limit by a factor of 1.5 before the crystal was inserted, and exceeded the limit by a factor of two after the crystal was inserted.

During that same store, the extraction rate was also studied as a function of luminosity. Only 6 of the proton bunches were colliding with antiprotons. Colliding and non-colliding proton bunches were observed during the same counting interval by employing two gates triggered on different bunches. With a typical bunch luminosity of $0.4\cdot10^{30}$ cm$^{-2}$s$^{-1}$ and typical circulating proton and anti-proton bunch intensities of 4 to $6\cdot10^{10}$ the extracted beam rate increased by factors of 4 to 8 for proton bunches that were colliding. The extraction rates resulting from collisions at CDF and at D0 were about the same. Further discussion is found in Section X below.

## VI. EXTRACTION EFFICIENCY

The principal motivation of this experiment was to explore the possibility of producing useful extracted beams during collider operation without undue impact on collider experiments. To achieve such beams it is desirable for the efficiency to be as high as possible. Efficiencies of up to 15.4% were measured in a recent CERN 120 GeV experiment[10,11] and 18% at 270 GeV[12]. Recently a group at IHEP, Russia, working with an extremely short crystal have achieved efficiencies of the order of 45%[13].

"Efficiency" in this context is defined in two ways. One practical definition, which we call the "extraction efficiency", is the rate of beam extracted into the extraction channel divided by the increase in the total circulating beam loss rate in the accelerator after the crystal was inserted. From an accelerator operation point of view, this is a practical definition, for it is the ratio of protons extracted to protons lost from the accelerator resulting from the crystal insertion. This definition was used in the CERN experiment.

Several effects contributed to this efficiency. The major contribution to lowering this efficiency was from protons which interacted with the crystal (12.9% of an interaction length) on one of their several passes through the crystal. A second contribution was from protons which were initially channeled, but then dechanneled after being bent through approximately 50 to 300 μrad. In this angular range, a proton is within neither the extraction channel aperture nor the circulating beam aperture and is intercepted by the Lambertson magnet steel or a collimator. A third contribution was from protons which were fully channeled but left the crystal through the beam-side surface of the crystal because they had a large negative horizontal angle, referred to hereafter as the "surface loss" contribution. On the other hand, unchanneled particles which only multiple scatter in the crystal rarely leave the aperture of the circulating beam, since the rms multiple scattering angle is 10.8 μrad, comparable to the angular divergence of the circulating beam.

While the numerator (the rate of beam extracted) was straight forward to measure, determining the change in the total loss rate from the accelerator was usually difficult. The loss rate was determined by doing a least squares fit to the slope of the circulating beam intensity versus time for several half-hour periods before and after the crystal was inserted. The error in this slope was determined from the rms deviation of the data from the fitted line. The variation with time of the loss rates before the crystal was inserted, resulting from various instabilities in the accelerator, usually exceeded the difference between the crystal-out and crystal-in loss rates.

No successful measurements of this type were obtained, but the data can be used to set 90% confidence level lower limits on the extraction efficiency (see Table II). The values varied considerably from session to session, a result of the fact that while the upper limit in the change of the loss rate was quite similar in all sessions, the extraction rate varied considerably.

TABLE II. Measured lower limits for the extraction effieciency. In column 2, "P" means "proton-only" and "L" means "luminosity driven."

| Session number | Type of store | 90% CL lower limit for extraction efficiency |
|---|---|---|
| 16 | P | 0.293 |
| 18 | P | 0.310 |
| 20 | L | 0.041 |
| 24 | L | 0.048 |
| 25 | L | 0.053 |
| 30 | P | 0.117 |

A second way to measure the efficiency was to compare the number of protons that interact with the crystal when its vertical angle is not aligned to the beam with the number that interact when it is correctly aligned for maximum channeling. Fewer interactions are observed when the crystal is well aligned with the beam because fewer protons interact with nuclei when a significant fraction of the beam incident on the crystal is channeled. This is called the "channeling efficiency" and is defined as the difference between the aligned and unaligned interaction counter rate, divided by the unaligned rate.

The "surface loss" mentioned above does not lower this efficiency, and the dechanneling losses contribute only partially (once a proton has dechanneled after channeling through part of the crystal, it has less than 12.9% probability of a nuclear interaction). Thus this efficiency is expected to be slightly higher than the extraction efficiency.

A simple model for the ratio of these efficiencies can be constructed. Of the total number of particles, $N_{tot}$, incident on the crystal (sometimes several times because of the multi-pass phenomenon), their ultimate fate falls into three categories: those extracted, $N_{ext}$; those interacting with the crystal, $N_{int}$; and those intercepted by an accelerator collimator, $N_{col}$, after either dechanneling or large amounts of multiple scattering. When the crystal is very misaligned from the optimum channeling angle, all the particles interact with the crystal (ignoring for the moment the possibility that particles multiple scattered in the crystal are intercepted by a collimator).

Using the above definitions of extraction efficiency, EE, and channeling efficiency, CE, it is straightforward to show, CC = EE(1 + $N_{col}/N_{ext}$). One then needs a model for the factor (1 + $N_{col}/N_{ext}$). To first order, it is expected that those which centripetally dechanneled (which occurs entirely in the first 64 µrad of bending) or did not channel will remain in the aperture of the accelerator and reencounter the crystal on a later turn. However, those which dechannel in the angular range 60 to 580 µrad are expected to intercept a collimator. It is observed that the dechanneling tail is 20% of the extracted beam in the angular range 290 to 590 µrad. Extrapolating linearly (not necessarily a correct assumption) to the angular range 60 to 580 µrad, a value 1.35 for the factor (1 + $N_{col}/N_{ext}$) is obtained.

In order to do a better job of estimating the real effects of the collimators, a fast Monte Carlo program was written which simulated both types of dechanneling, calculated the probability of an interaction in the crystal for unchanneled and dechanneled particles, and added Gaussianly distributed multiple scattering angles for those which did not interact. These particles were then traced to the three critical collimators used to scrape the beam during collider mode. The above factor was found to vary from 1.2 to 1.4, depending on the setting of the collimators and the model used to extrapolate the angular distribution of the normal dechanneling to the region where the tail was invisible. In this model the number of particles intercepted by collimators after multiple scattering in the crystal when the crystal was not aligned to the optimum channeling angle was not negligible (typically 15%).

In operation, the interaction counters were also sensitive to fluctuations arising from such effects as small horizontal deviations of the circulating beam. Some of these effects could change in an unpredictable way in the time it took to do a typical scan. For example, Fig. 12 shows both the extracted beam and the interaction counter rate as a function of $\Theta_v$. The time ordering of the points in the scan is from right to left. Note that the interaction rate is considerably higher off-peak to the right than to the left, and is falling with time. In this case the time dependence is understood: the scan was begun immediately after the crystal had been inserted to 5 mm from the beam, and it takes several minutes for the rate of beam incident on the crystal to decrease to an equilibrium rate.

To mitigate such a possible time dependence, the best measurements were obtained by moving the crystal quickly back and forth from an aligned to a very unaligned vertical angle. An example of such data is shown in Carrigan *et al.*[6] In all such measurements the crystal was moved quickly back and forth between three positions several times: aligned for maximum channeling, very unaligned at an upward angle, and very unaligned at a downward angle. The interaction rate was averaged at the three angles, and errors determined from the rms deviation of the data from the mean. A weighted average was taken of the two off-peak values for use in the calculation of the channeling efficiency.

A background was also subtracted from the measured interaction rate. Experience showed that as the crystal was moved from its fully retracted position, the interaction rate increased as soon as the crystal was within the aperture of the accelerator, long before any crystal extraction was observed, and remained constant until the onset of extraction. This background was probably the result of a low intensity, diffuse halo intercepting the large mass of the crystal holder. This background varied from 6% to 36% of the off-peak interaction rate, depending on the session, but was quite constant during a session. An error estimate for the background was made and propagated through the calculation of the efficiency.

The channeling efficiencies resulting from these measurements are shown in Table III. The efficiency need not be the same from session to session (for example, the collimator settings were different in each session, and distance of the crystal from the beam differed), but should be the same for different on/off scans during the same session. Therefore, weighted averages are shown in Table III for sessions 28 and 30. In fact, the weighted averages of the three sessions are remarkably similar, despite the fact that one of the three sessions was a proton-only fill in which the collimators were out of the beam.

TABLE III. Measured channeling efficiency for various sessions using "on/off" data. Column 2 has the same meaning as in Table II. For two of the session, the on/off data was taken more than once at different times during the session, and the weighted average (wt. ave.) for the session is also presented.

| Session number | Type of store | Channeling efficiency | Error |
|---|---|---|---|
| 27 | L | 0.244 | 0.080 |
| 28 | L | -0.055 | 0.203 |
| 28 | L | 0.411 | 0.235 |
| 28 | L | 0.566 | 0.159 |
| 28 | wt. ave. | 0.348 | 0.111 |
| 30 | P | 0.321 | 0.112 |
| 30 | P | 0.332 | 0.154 |
| 30 | wt. ave. | 0.325 | 0.091 |

The simulation[3] predicted an extraction efficiency of 35% for a realistic crystal. The same simulation program gives a value consistent with the efficiency measured at 120 GeV at CERN[14] Using the factor 1.3 mentioned above relating the extraction and channeling efficiencies, our typical measured channeling efficiency of ~30% leads to an extraction efficiency of ~23%.

One can also calculate channeling efficiencies using the exact same methodology with the data from the $\Theta_V$ scans, but a choice must be made about which points to call "on-peak" and "off peak". This was done using the plot of the extracted beam versus $\Theta_V$. This approach has been discussed in Ramachandran's thesis[5]. These results are not as credible as those of Table III, for in any of these scans, there may have been an unknown time-dependence in the interaction rate resulting from other changes in the accelerator.

It would have also been interesting to measure the extraction efficiency in kick mode. In this case the denominator (the number of protons lost from the circulating beam after each kick) was easy to measure, for in these proton-only fills, the kicks were nearly the only source of loss, and amounted to typically $2 \cdot 10^9$ protons/kick. However, it was very difficult to measure the numerator, the number of protons extracted following each kick. This intensity range, $\sim 10^9$ protons/kick, was too low for the toroid or SWIC to function. Attempts to calibrate the pulse height of the counters, operating at reduced voltage in "calorimeter mode", and to calibrate the CCD using aborts of the entire circulating beam were not successful because of the difficulty in injecting and measuring such small amounts of circulating beam.

## VII. TIME DEPENDENT AND BEAM HALO EFFECTS

A wide variety of time-dependent and beam halo effects were observed in the course of E853. Characteristic time constants ranged from tens of microseconds to many seconds. Some were inextricably related to the nature of channeling such as multi-turn processes where some fraction of the beam on an incorrectly aligned crystal scattered to an aligned angle. Others like the behavior of the extracted beam rate as a function of crystal motion in and out of the beam were more directly related to accelerator beam properties.

As discussed earlier, data was taken in two modes, kick and diffusion. In the kick mode an individual bucket was kicked on to the crystal. In the diffusion mode some of the beam halo diffused out to larger distances because of such effects as beam gas scattering and magnet ripple. Two special cases, scattering by a flying wire and RF noise, are discussed in later sections.

The following sub-sections discuss the kick mode behavior, beam halo effects observed through crystal and collimator motion, and oscillations and modulations.

### A. Kick mode

The major time structure in kick mode was due to accelerator beam dynamics. Because of the accelerator phase advance between the kicker magnet at E17 and the crystal at B48, on the first turn following the kick the beam had moved away from the crystal (illustrated in Fig. 3 of Ref. [4]). On turn 2, and again on turn 7, the beam was at maximum amplitude towards the crystal. Sizeable extraction was expected on turns 2 and 7. Extraction occurred on the turns in between only when beam that was not channeled in turn 2 multiple scattered to a different vertical angle and then returned to encounter the crystal with the correct angle on a later turn.

The in-between turns are called "wrong side" turns, in contrast with "right side" turns such as 2 and 7. This process has been successfully modeled in simulations prepared by Biryukov[3] and Bogacz et al.[2] As would be expected, the detailed microstructure of the turn-to-turn variation was quite sensitive to the accelerator tune. This was also observed experimentally for tune changes in different runs. Indeed, information on the micro-structure was more indicative of tune changes than underlying physics related to channeling.

Fig. 13 illustrates the behavior for the actual data and the Bogacz simulation for the first 23 turns (about 400 µs). In this time domain the extraction data rate based on the pulse height in a scintillator is reasonably modeled by the Bogacz simulation program except that extraction persists longer than predicted. Not surprisingly, the detailed pattern from the Bogacz simulation is tune-sensitive.

Fig. 14 shows the extraction pattern for a longer period (200 ms or 10000 turns) for the large, right side turns. The measured initial total decay time ranges from 0.6 to 2.5 ms. In that period the signal decreases by 10-60% for different kicks. Short values for this decay are more in line with the Bogacz simulation. It is difficult to produce 1/e times of less than 0.9 ms with the Biryukov simulation. Under somewhat artificial conditions the Biryukov simulation can generate time constants greater than 2.0 ms. Changing the interaction length and the horizontal kick by substantial amounts produces less than a 20% change. Changing the horizontal misalignment by 10 mrad can double the decay time.

A related issue is how long it took to reach equilibrium when the crystal was vertically misaligned. Multiple scattering resulting from multiple passes through the crystal is the underlying mechanism for this effect. Fig. 15 shows the time distribution for 38 turns after a kick for a -60 µrad misalignment. Data sets were also obtained for -40 and -20 µrad misalignments.

The highest "right side" turn data (circle with an x) has been fitted with an exponential form. The form (dictated by available fitting routines) also effectively incorporates a time before any beam is extracted. At -60 µrad no beam appeared for several turns. The rise time was 170 µs or 8.5 turns. The asymptotic extraction signal was about 70% of the on-peak case. At -40 µrad, the initial delay was still several turns and the time constant was 120 µs. The asymptotic signal reached the on-peak value.

A naive picture gives some insight into the process. On every pass through the crystal (approximately every other turn) the particle was scattered 10 µrad by multiple scattering. To change by 60 µrad required on average $6^2$ passes or 72 turns to full equilibrium while 30 turns were actually needed to get to 90% of the asymptotic value. Part of this difference was probably due to the effect of the beam divergence which was already 14 µrad after 6 kicks and continued to grow with each successive kick. With beam divergence and multiple scattering the time constant should go as $\Theta_m^n$ where $1 < n < 2$ and $\Theta_m$ is the misalignment angle. The Biryukov simulation reproduces the general features of the data but the simulation time constants are roughly twice as long (200 µs at 40 µrad and 400 µs at 60 µrad) and the pattern is not as smooth as the data. A wider effective accelerator angular divergence than used in the simulation might have led to shorter times.

Another related subject is the time for a "wrong side" turn to come out at the same rate as a "right side" turn. This equilibration is probably a result of non-linearities in the accelerator lattice. Observationally the amplitude of wrong side turns remained small for 20 turns and then increased and came into equilibrium with the right side turns with a time constant comparable to the initial fall time for right-side buckets. This is illustrated in Fig. 16. Neither of the simulations reproduces this feature since non-linearities were not included. Over several hundred turns in the Biryukov simulation the wrong-side turns remain small with no discernible pattern. The experimental data has been fitted with an exponential distribution coming into equilibrium with an asymptote. An effective initial delay before a signal of about 350 µs (18 turns) was followed by a rise with a time constant of 520 µs.

A second interesting feature appears in Fig. 14. This is a 10 ms period where the extracted rate slowly changed. The envelope decreased slowly at first and then more quickly. The initial slow portion extended for on the order of 20 ms. The time constant to the 50% point was 30 to 50 ms. The data hints that this might have been mixed in with some sort of 30 to 50 ms peak-to-peak oscillation (see later discussions under oscillations). Nether of the simulations shows this effect. Again, a plausible suggestion for a driving mechanism is a non-linear effect in the accelerator.

There were also effects extending out to many seconds after a kick. These were observed using video recordings at an effective rate of 60 frames a second with a CCD camera monitoring the scintillator screen in the extracted beam line. Three types of beam spot pictures were observed. One was the first frame corresponding to the first 16 ms of Fig. 14. This was typically followed by several frames of decreasing intensity corresponding to the decay portion of Fig. 14. A third set of pulses was observed at times well separated from the initial kick interspersed with many blank frames. Since the CCD-scintillator system was non-linear it was not possible to make more than a qualitative judgment of the pulse intensity.

The number of these pulses versus time summed over several kicks is shown in Fig. 17 for all pulses (X's) and for the lowest intensity ones (open circles). Note that this distribution is for the number of separated pulses, NOT the number of particles down the beam line. This distribution measures some convolution of the effect of small perturbations long after the kick with available halo generated by the kick that has not yet been scraped. The distribution dies with a time constant of 0.8 s. After five seconds there is an almost flat spectrum of very small pulses.

### B. Beam halo effects

Several E853 observations relate to the character of the circulating accelerator beam halo. The effect of retracting the crystal or a collimator a small distance sheds light on the diffusion rate of the halo. This rate is related to the beam halo phase space density and non-linear effects in the accelerator system. Likewise, the number of kicks required to move beam out to the crystal or "grow" the beam is an indicator of the halo density. Moving the crystal or a collimator generated related information. When the crystal was moved in there was an initial quick beam rise that died over a minute or so followed by a sustained rate that could persist for hours.

Unperturbed beam lifetimes in the Tevatron are quite long. For E853 typical proton beam lifetimes ranged from 70 to 90 hours, while the luminosity lifetimes were 11 to 18 hours. Because diffusion rates were long and data collection time was short, little data was taken where the incremental retraction of the crystal was small enough to produce a noticeable effect and the time interval before the next disturbance was long enough. Fig. 18 shows the best information available from a diffusion run in collider mode. After a 200 μm retraction the count rate dropped by a factor of 4 and then increased with a time constant of 2.2 minutes. Weaker information for a 50 μm retraction shows an initial drop of a factor of 2 followed by a rise with a time constant of ~15 s. Based on this information it is difficult to speculate on whether the time constant is linear or goes with the square of the retraction distance.

Collimator effects were also studied. The effect of the positioning of the three scraper collimators at D17 and A0 used to protect the collider experiments from beam halo was clearly interwoven with crystal position. When the crystal was effectively closer to the beam than the collimators the situation was different than when the crystal was shadowed by the collimators and diffusing beam was mostly lost on the collimators. For example a 5 mm retraction of the crystal (well outside the collimators) lowered the rate precipitously, and even after 20 minutes there was no sign of increase.

In three study sessions the collimators were retracted almost simultaneously in small steps (typically 2 mils) near the end of the runs by total amounts that varied from 0.5 to 1.2 mm. In all cases the extraction rate rose, but the rate of rise as a function of collimator position varied. This may be a result of the fact that the collimators did not have the same initial settings in the three runs. In a study with a total motion of 0.5 mm and the crystal 5.5 mm from the beam center the normalized extraction rate of rise was a factor of 5 per mm of collimator opening (normalized to the rate at the initial collimator positions) while in two later sessions with crystal-beam separations of 5.3 and 5.5 mm the rate of rise per mm was in the 1-1.5 per mm range. A plot of the extraction rate as a function of collimator position for one of the latter sessions is shown in Fig 19. The data is satisfactorily represented with a linear fit. In the three sessions the D0 proton loss rate rose by 5% to 20% as the collimators were opened.

Studies of the time to reach equilibrium after a collimator move were complicated by relatively quick collimator changes with few measurements taken between adjustments. The time required to adjust the three collimators (on the order of 10 s) was also a limitation. Fast time plots and counting rate information indicated an initial fast rise of the count rate in less than 30 s, and perhaps much less. Information is available on only one relatively long quiescent period after a collimator move. Twenty minutes after completing a collimator moving session that had lasted 13 minutes the rate had risen by 30% (four times the estimated standard deviation of the measurement).

Somewhat related information is provided by measurements of the number of kicks required to move the edge of the beam out to the crystal so that the asymptotic rate was reached. Typically it took 5-7 kicks to grow the beam. Fig. 20 shows the behavior as a function of kick number for a typical case when the crystal to beam center distance was 3.54 mm. The asymptotic rate was reached in 5 kicks. In a case with the crystal 4.5 mm from the beam it took 10 kicks to reach the asymptotic rate.

Another related parameter is the time to reach an equilibrium extraction rate after the crystal was moved in. Normally there was only a short time interval available for analysis after the crystal was moved (typically 1-2 minutes) before some other change was made. Since equilibrium times were short, only data gathered starting a few seconds after a move could be used. A measure for the relaxation rate was obtained by fitting a linear slope to the extraction rate as a function of time to get $-(1/R)dR/dt$ which is equal to $1/T_e$ ($T_e$ is the time to approach equilibrium and R is the extraction rate). A typical value of $1/T_e$ was $0.15\pm0.03$ min$^{-1}$. This means the rate dropped by 15% in a minute. The error bar is based on the uncertainty in clock time for the measurements (0.33 min). The value of $1/T_e$ at various distances from the beam and in various study sessions varied from 0.03 to 0.4 min$^{-1}$. The equilibrium time seemed to shorten as the goniometer was moved in.

After the initial rapid fall discussed above the rate decreased with a time constant (1/e) between 0.5 and 5 hours under a variety of operating conditions. Typically there is no statistical difference between a linear and an exponential fit to these long-term halo decays.

### C. Oscillations and modulations

In the course of E853 several different phenomena appeared that were initially ascribed to some sort of oscillatory behavior. The possibility of oscillations with periods on the order of milliseconds was investigated and none were found.

The possibility of extracted beam modulation resulting from the Main Ring was also investigated. Initially there was concern that magnetic fields from the Main Ring or its power busses would produce Tevatron orbit perturbations large enough to modulate the extraction rate. The possibility of Main Ring oscillations was studied in kick mode by looking at the number of pulses seen in the CCD camera long after the kick as a function of time in the Main Ring cycle. Fig. 21 shows this distribution summed over the Main Ring cycle for several kicks. No statistically significant peaking in the distribution of the pulses was observed. Coincidence counter data between air gap 1 and air gap 2 also gave little or no indication of Main Ring modulation.

One oscillation-like behavior was observed in kick mode studies. It is illustrated in Fig. 14. This behavior showed wide variation from kick to kick. For example, the period between peaks varied from 25 to 75 ms. These oscillations were overlaid on the exponential decay of the signal after a kick. Only a few maxima were observed because the scope trace was only 200 ms long. As a result it was difficult to draw a conclusion. If this was an oscillatory effect it might have been related to synchrotron oscillations with a characteristic frequency of 39 Hz at 900 GeV or a sub-harmonic of the 60 Hz power.

## VIII. RF-DRIVEN EXTRACTION

The impact of a transverse RF noise signal generated by a horizontal damper located at F11 was tested during two sessions. The effect of turning the RF on or off was almost immediate and had a significant effect on the extracted beam. It was so significant that these tests were complicated by saturation problems in the larger principal counters even in a special proton-only store with 84 rather than 6 circulating bunches.

The smaller finger counters showed little or no evidence of saturation and also had a low background rate with noise off. Fig. 22 illustrates the behavior of the crossed finger counter coincidence with increasing rms RF voltage for a crystal to beam separation of 3.94 mm. The curve shows a fit to the data of the form $R = 15.2 \cdot V_{rf}^2 + 22$ where R is the count rate. A $V^2$ dependence is expected from an analysis of the equations of motion in action-angle variables[15].

The possibility that accidental coincidences contributed to the crossed finger rate was studied and was found to contribute 1.5% at a low $V_{rf}$. This percentage should increase as $V_{rf}^2$ and was 12% at the highest voltage shown on Fig. 22. Accidentals were subtracted from Fig. 22.

In spite of the saturation problems it is still possible to make interesting estimates of upper limits on the time constants for RF changes. When the RF signal increased the count rates (typically taken several times a minute) rose immediately, suggesting the rise time was less than a minute. Fast time plot information on rate meters rose within a 10 s period and probably somewhat more rapidly, suggesting a time constant (1/e) of 2-3 s for $V_{rf}$ = 4.5 V on the RF damper. This voltage should have resulted in a scattering angle of $\Theta = 4 \cdot 10^{-4}$ μrad and a jump at the crystal of 0.04 μm/turn. If this was a linear process with a linear growth with time it would have taken 1.5 s for a particle to move out to 5 $\sigma_x$. This naive calculation is meant to show only order of magnitude agreement with the observations, for the correct model for this stochastic process is complicated. Furthermore, protons already close to the crystal when the noise was turned on would reach the crystal faster.

The fast time plots also provided information on the effect of turning off the RF damper or turning the RF down. When the damper was turned off from $V_{rf}$ = 4.5 V the extracted beam rate decayed over a 15 s time period with a time constant (1/e) of about 10 s.

## IX. EXTRACTION DRIVEN WITH A FIBER SCATTER

Asseev[16] and others have suggested that an amorphous scatterer placed closer to a circulating beam than an extraction crystal might facilitate bent crystal extraction. During E853 an investigation directed toward this possibility was carried out using the Tevatron "flying wire" system located at E11 and E17. The results were both interesting and curious.

During collider operations at the Tevatron three 30 µm carbon fibers called flying wires are rotated through the circulating beam at a velocity of 5 m/s about once every half hour. Each fiber passes through the beam twice. These are normally used to measure the profile in x and y and the momentum spread of the circulating beam. The "prompt" time distribution of the particles scattered off the fibers is related to the beam shape at the time the wires fly. For a beam width of $\sigma$ = 0.7 mm at the wires the full width at half maximum of the counting rate distribution as a function of time should be 330 µs.

The flying wires were carefully monitored during diffusion mode running in the latter part of E853. A digital sampling scope operating in the peak detector mode tracked pulse heights for both an interaction and an extracted beam scintillator. An individual count corresponded to 0.1 volts based on observations during setup (the discriminators had been set at 0.03 V) and baseline analysis. The wire signals were clearly visible in the extracted beam so that count rates taken during wire flys had to be removed from the data analysis. The flying wires were also seen at CDF and D0 where the detectors were gated off when the wires flew.

The voltages on these scintillators had to be kept high since they were also being used for counting. As a result there was a significant possibility of saturation and photomultiplier power supply drain because of the instantaneous rates. The phototube bases were equipped with transistors to provide some protection against slumping voltages. Post-run analysis indicated that the extraction scintillator probably should have been able to sustain the rates. However in view of the unusual features exhibited by the extraction scintillator this potential problem should be kept in mind.

One reason to believe that there were no significant saturation effects or slumping of the phototube voltages is that the scintillators did show the characteristic bunch structure. A second way to evaluate the possibility of saturation is to look at the ratio of prompt extracted and interaction signals during regular and flying wire running. During normal running the interaction detector rate was 10 times smaller than the extracted signal so that the interaction detector should have saturated less. For a typical wire fly it was 17, in the opposite direction expected for saturation.

Protons in the beam interact with a carbon fiber in one of two ways. They can undergo nuclear collisions with characteristic scattering angles of 300 µrad or experience atomic multiple scattering with a scattering angle of 0.16 µrad. The chance of a particle passing through the wire twice is about 1/4. Since the multiple scattering was small relative to the beam divergence it probably resulted in little beam growth. For $10^{12}$ circulating protons on the order of $10^7$ undergo a nuclear interaction with a wire. Of these on the order of 3% or $3 \cdot 10^5$ per fly have scattering angles small enough to remain within the accelerator and strike apertures far downstream like the crystal. For a typical fly $0.7 \cdot 10^5$ of the particles scattered by a wire were extracted by the crystal. For an extraction efficiency of 30% this suggests that on the order of $2 \cdot 10^5$ were striking the crystal corresponding to a large fraction of the losses around the ring. This is not inconsistent with the fact that losses were observed elsewhere such as at the two collider detectors.

A distinct and puzzling feature of the flying wire data for E853 was that the extracted beam detector for the aligned crystal showed a delayed signal that started some milliseconds after the prompt signal and extended for tens of milliseconds (Fig. 23). The overall length of the time distribution is reminiscent of the pattern from kick mode, reproduced in Fig. 23 from Fig. 14. For a 60 µrad misalignment the size of the prompt extracted signal was consistent with a normal multi-turn alignment distribution. That is, the rate was about a factor of eight smaller than for an aligned crystal. However there was no significant delayed extraction signal in the misaligned case. This may indicate that the delayed peak was an artifact of the higher intensity when the crystal was well aligned. For the aligned condition the total delayed extracted yield was about six times larger than the prompt yield. The interaction signal also had a delayed signal (see shoulder in Fig. 23 but with a shorter duration. The integrated delayed interaction signal was two times larger than the prompt interaction signal.

The observed behavior could be explained by postulating that the prompt extracted and interaction signals from a wire were due to wide angle particles from nuclear interactions in the wire striking the crystal. The delayed signals could have been due to multipass effects that continued over 500-1000 turns. As time went on halo evolution might have favored the preservation of smaller angle particles that still remained tightly collimated relative to the circulating beam so that the extraction signal dominated. However it seems implausible that this effect could have resulted in such a huge dip and peak. Perhaps the most plausible explanation is that it was caused by photomultiplier saturation and slumping voltage followed by a recovery.

The prompt rise time was 400 µs (from 10-90% of peak rate) for both the extraction and interaction signals (see Fig. 24). If this had been Gaussian, it indicates a sigma for the beam of 240 µs or 1.2 mm. While this is somewhat larger than the actual beam sigma at E11 it is not unexpected in view of various sources of broadening. However the FWHM for the extraction counter distribution is more like 1 to 1.1 ms, implying a σ of 470 µs. This could be reasonably interpreted as multi-turn broadening. Such broadening is also present in the interaction signal but can not be estimated easily. The extraction signal 60 µrad out of alignment also has a similar rise time but is narrower and more Gaussian-like suggesting that indeed multi-turn effects may be responsible for broadening the prompt, aligned extraction signal.

The falling side of the interaction rate distribution initially looks like a Gaussian but soon develops a long tail. The distribution has three exponential parts (see Fig. 25); 1) in the interval 1 to 2 ms after the peak the decay time is 5 ms, 2) in the interval 2 to 4 ms there is a faster decay with a time constant of 2 ms, and 3) in the interval from 5 to 20 ms after the peak the decay time is 6 ms. These regions are quite distinct. It is possible that this decay distribution is in some way reflecting the actual nuclear interaction angular distribution with the longer times mirroring partial contributions from very small angle scatters that could require more turns to reach the crystal.

The rise and fall times for the delayed extraction signal are hard to characterize. They range from 12 to 26 ms with the longest ones occurring for the wire used to measure momentum spread. The times for the second fly in each of the three wires are 20-25% longer than the first fly. On the other hand there is no indication from the prompt signal that the core beam diameter had grown after the first fly. The x and y distributions are about the same. For both the x and y wires the rise in the delayed extraction signal was approximately linear. The decay time ranges from 2 to 5 ms. This decay time is difficult to characterize because of other effects such as modulations.

Several features of the flying wire study are worth emphasizing. First there was a significant prompt extraction that occurred with a plausible efficiency. There was also an even stronger delayed signal that may have come from multipass effects. The features of this data may deserve further analysis and simulation. For this reason details of the data have been presented although concerns about scintillator counter saturation certainly cloud interpretation.

**X. LUMINOSITY-DRIVEN EXTRACTION**

During the first E853 parasitic collider session the channeled flux was substantially higher than earlier dedicated runs with the crystal at the same relative distance from the beam. A principal candidate for a driving mechanism was the effect of interactions at the two luminous regions for the two collider experiments on the emittance of a proton bunch. In one picture of this process the effective scattering target is the antiproton current times the geometrical part of the luminosity (see below) so that the extracted beam rate is proportional to the luminosity L. The luminosity is equal to a geometrical factor times the product $I_p \cdot I_a$ where $I_a$ and $I_p$ are the proton and antiproton currents respectively.

Since several factors influenced the extracted beam rate it was interesting to isolate luminosity from other effects. The equivalent in a fixed target experiment would be a target in-target out measurement. Several approaches were considered; 1) changing the arrival times of the proton and antiproton bunches at the interaction points so they did not collide (cogging), 2) displacing the antiproton beam at the interactions points, and 3) eliminating some of the antiproton bunches and thereby removing the "target". All of these required semi-dedicated running. Cogging was tried several times and led to ambiguous results because the proton beam invariably moved transversely at some point during the exercise due to lattice dispersion. No opportunity arose to try displacing the antiproton beam, but it might not have worked for the same reason cogging did not work. The approach which worked was to run with a special store with 36 proton bunches of which only six were colliding with three antiproton bunches.

The test of the luminosity-driven process was to first check that the beam extraction rate was substantially higher with a gate timed to overlap a colliding bunch than it was for a non-colliding bunch. Then the relationship of the count rate to the bunch luminosity was investigated.

The Tevatron operated at 900 GeV with the separators on for this store. The 36 proton bunches were spaced in groups of twelve distributed evenly 1/3 around the ring. The individual proton bunches in the groups of 12 were each separated by 395 ns. Three bunches of antiprotons were circulating spaced as though they were in one group of twelve with positions A1, A5, and A12 filled. This resulted in 6 sets of bunches colliding at the interaction regions shown at the top of the columns in Table IV.

TABLE IV: Bunches colliding at B0 (CDF) and D0. The notation P36•A12 means that proton bunch 36 was colliding with antiproton bunch 12.

| B0 (CDF) | D0 |
|---|---|
| P13•A1 | P25•A1 |
| P17•A5 | P29•A5 |
| P24•A12 | P36•A12 |

Data were recorded for the number of protons and antiprotons in each bunch as well as the transverse size of the bunches. Typically the number of protons in a bunch was $50\text{-}70 \cdot 10^9$ and the number of anti-protons was $40 \cdot 10^9$. Information was also available on the population numbers for nearby satellite bunches. Some D0 luminosity information was also recorded. There were some problems in recording information on luminosities and several other ACNET variables because the complicated 36 on 3 collision pattern resulted in the need to reset many ACNET timing signals. As a result some information on luminosity was lost.

With the exception of P25 the transverse bunch widths ($\sigma$) were all on the order of 0.5 mm (at the flying wire location) for this run. For P25 the bunch width was 0.65 mm. No data was available on the horizontal anti-proton bunch width.

Two sets of measurements were taken. In the first set an effective gate consisted of three windows each of 470 ns spaced evenly over a turn. Two such independent gates were used. The gates were broader than the spacing between two bunches because it was sometimes difficult to set the Tevatron timing signal to better than 100 ns. (In several cases the gates were set to nominally non-colliding time values that were later shown to collect colliding bunches.) Typically one of the two gates was set on a colliding bunch and the other on a non-colliding bunch. With this arrangement collisions were detected that resulted from both the B0 and D0 interaction regions (set 1). Halfway through the measurements a different arrangement was adopted with only one window (rather than three equally spaced ones) to see the individual effects of D0 and CDF (set 2).

During the course of these two sets of measurements the ungated count rates dropped steadily. Both the colliding and the non-colliding bunch rates exhibited this behavior. This suggests that the count rate drop was not an effect related to luminosity. As a result of this drop in rate it was necessary to move the crystal 300 μm further in toward the beam at the end of the first set (from a crystal to beam separation of 5.58 mm initially) to increase the instantaneous count rate at that time by a factor of 1.82. The time dependence for the first set of measurements was exponential with a lifetime of 1.5 hours (1/e). This is shorter than either the luminosity lifetime (11 hours for D0) or the proton beam lifetime (70 hours). It may have corresponded to some sort of halo lifetime. The time dependence was folded in for the analysis.

The effects of leading and trailing satellite buckets were included but additional small terms associated with crossings upstream and downstream of the colliding points were not. The corrections to just using the product of the main bunch populations were less than 3%. The calculated bunch luminosity was then:

$$L_{pa} = KI_p I_a \frac{A}{A_p A_a} = KI_p I_a L_{geo} \quad (X\text{-}1)$$

where K is a constant to normalize to the measured luminosities, $A_p$ is the area of the proton beam (assumed to be $\pi\pi\sigma_{pv}\sigma_{ph}$) and $A_a$ is the area of the antiproton beam. A is the overlap area and is calculated using the smallest dimensions of either of the two beams. $L_{geo}$ is the geometric part of the luminosity so that $L_{geo} = A/A_p A_a$ in $mm^{-2}$.

The bunch luminosities were also measured for the D0 cases. These measured luminosities, $L_m$, are given in Table V in units of $10^{-30}$ $cm^{-2}$ $s^{-1}$. Values for $L_{geo}I_p I_a$ were found by multiplying the bunch populations (given in units of $10^9$ particles) in Table V by the relevant beam sizes. K was determined by averaging the three ratios of the measured luminosities divided by $L_a I_p I_a$ to give $K = 1.338 \cdot 10^{-4}$. The calculated bunch luminosities ($L_{pa}$) are also given in Table V. The measured and calculated D0 luminosities (bunches P25•A1, P29•A5, P36•A12) agree quite well. The bunch with the largest difference between the calculated and measured luminosities was P25•A1 where the proton beam dimensions were 30% larger and the current was down by 45%. Note that there was a wide spread in the luminosities of the different bunches.

Table V. Calculated ($L_{pa}$) and measured ($L_m$) bunch luminosities, in units of $10^{30}$ $cm^{-2}s^{-1}$. $I_p$ and $I_a$ are the measured number of protons and antiprotons in the main colliding bunches in units of $10^9$. The product $I_p I_a$ includes the small contributions from satellite bunches.

| Colliding Bunches | $I_p$ | $I_a$ | $I_p I_a$ | $L_{geo}$ | $L_{pa}$ | $L_m$ |
|---|---|---|---|---|---|---|
| P13•A1 | 54.6 | 42.5 | 2364 | 1.445 | 0.457 | |
| P17•A5 | 45.1 | 41.9 | 1934 | 1.308 | 0.339 | |
| P24•A12 | 58.6 | 44.6 | 2636 | 1.238 | 0.437 | |
| P25•A1 | 29.3 | 42.5 | 1273 | 0.713 | 0.122 | 0.132 |
| P29•A5 | 70.9 | 41.9 | 2995 | 1.330 | 0.534 | 0.494 |
| P36•A12 | 39.6 | 44.6 | 1822 | 1.286 | 0.313 | 0.309 |

The count rate difference between colliding and non-colliding buckets is illustrated in Fig. 26. The measured count rates are compared to calculated rates based on luminosity plus an averaged background (see next paragraphs). Measured count rates are shown as histograms with the widths of the gates. Colliding cases are shown as solid histograms while proton-only cases are dashed. (Proton bunches P14 (set 1) and P26 (set 2) were nominally out-of-time but subsequently were found to overlap colliding buckets.) Since set 1 used a gate that covered three buckets, two of which were colliding, the rates for set 1 were roughly twice as large as set 2. The difference between colliding and non-colliding buckets is clearly visible in Fig. 26.

The time-dependent "background" was found by averaging the measured rates of the non-colliding bunches for set 1 ($B_{1av} = 4282/s$) and for set 2 ($B_{2av} = 2723/s$) extrapolated to an arbitrary $t_0$ (30 minutes before the start of the first set). The measured ratio for $B_1/B_2$ is 1.57 while the ratio was expected to be 1.65 because set 1 had three buckets within the gate rather than one for set 2, but the rate had been increased by 1.82 when the crystal was moved in for set 2.

To investigate rate dependence on luminosity the measured background-subtracted, time-normalized extraction rates ($t = t_0$) are plotted against $L_{pa}$ in Fig. 27. Measured values for set 2 are divided by 1.82 to account for the fact that the rates rose when the crystal moved in. A fit constrained to go through the origin gives:

$$C(/s) = (0.269 \pm 0.034) \times 10^{-25} \times L_{pa} \quad \text{(X-2)}$$

where $L_{pa}$ is in units of $cm^{-2}s^{-1}$ and the rate is based on the crystal to beam separation of 5.5 mm. All of the available information was used. The correlation coefficient is $r^2 = 0.55$. While this is not an impressive fit it does suggests that the expected count rate above background is proportional to luminosity. To the extent the data permits concrete analysis, the contribution of D0 and B0 to the extracted rate seem to have been the same. This formula was used to determine the calculated rates in Fig. 26. The appropriate luminosity was substituted and the background rate was added. Instantaneous values were used for the calculated rates (rather than rates extrapolated to $t_0$) so that count rates on the same bucket measured at two different times have different calculated values in Fig. 26. Set 2 cases were multiplied by 1.82.

Luminous bunches increased the rate substantially. Based on the fit in Fig. 27 (equation X-2) a net luminosity of $0.5 \cdot 10^{30}$ $cm^{-2}s^{-1}$ gave a luminosity on/off ratio of 4.1. The fitted measured count rate (less background) for that luminosity determined from equation X-2 is 13300/s. One way to quantify the luminosity on-off ratio is to ask how far the goniometer would have had to be moved in to match a non-luminous bunch to the original luminous rate. Increasing the non-luminous count rate by a factor of 4 at $x_g = 2.5$ mm was equivalent to decreasing $x_g$ by 0.4 mm or 0.7 $\sigma_x$.

It is also interesting to compare luminosity and RF-driven extraction. This can be done by comparing the ratio of the rate with the driver on to the rate with no driver under the same conditions. As noted in the last paragraph this ratio was 4.1 for $L_{pa} = 0.5 \cdot 10^{30}$ $cm^{-2}s^{-1}$. In the RF-driven case discussed in section VIII the rate for the finger counter coincidence is $R = 15.2 \cdot V_{rf}^2 + 22$. A ratio of 4 is obtained for $V_{rf} = 2.1$ V.

In summary, it is clear that luminous bunches produce substantially more extracted beam. A useful study would be an investigation of the actual physical mechanisms that influence the beam growth. It would be interesting to try to relate them to the beam loss rates. Models might be developed using emittance growth studies undertaken for the SSC and LHC.

## XI. APPLICATIONS

Multiple pass crystal extraction, first studied theoretically by Biryukov[17] and Taratin et al.[18] and since demonstrated in this experiment and at CERN[19], has made crystal extraction very interesting. With multiple passes, the non-channeled beam goes through the crystal many times until it interacts or channels and is extracted. The angular divergence of the extracted beam in the bend direction is small, on the order of the channeling critical angle.

The application of crystal extraction most frequently considered for TeV-range accelerators has been scraping of the halo in a collider to produce parasitic beams in the 1 MHz regime. In E853 a parasitic 120 Khz beam was extracted at C0 without undue background impact on the collider detectors. Parenthetically, the collider loss monitors were set conservatively during E853 and little head room was allowed for losses.

Crystal extraction can also work for dedicated extraction. While the efficiency is not as good as with an electrostatic septum, it can potentially be well above 50%. Recent Serpukhov studies[13] have found that short "O" shaped crystals with no straight ends and smaller radii of curvature increase the efficiency further, as expected.

With crystals the large bends relative to conventional electrostatic deflection offer intriguing possibilities for applications. For example, crystal extraction has been used with success in the tight lattice at Serpukhov. It could also be used to pack extraction facilities into oversubscribed straight sections at colliders. One ambitious suggestion is the possibility of directing neutrino beams toward cosmic ray facilities well off the plane of an accelerator[20]. The following subsections discuss TeV extraction and possible charm experiments including magnetic moment measurements.

### A. TeV Extraction at Fermilab

An interesting candidate for crystal extraction is a parasitic collider beam from the Tevatron A0 straight section into the fixed target experimental areas. A 1000 GeV beam would have 25% higher energy than any fixed target beam of the past. The beam could be useful for such projects as LHC development test beams. The spill structure would be uniform without the normal fixed target magnet ramping cycle. The beam-on time would correspond to the "on" time for the collider. The beam could be switched to several different experimental areas. Switching would lessen problems of operating the Switchyard above 800 GeV. The beam could also be operated at higher intensities for short periods in a dedicated mode.

For parasitic operation the crystal would operate like the E853 crystal. The principal difference is that a much larger bend angle would be required to avoid obstructions at A0 and to mate with the Switchyard beam line.

A logical location for a bent crystal in the Tevatron is just after the last A0 proton kicker magnet as shown in Fig. 28. This geometry has been discussed in more detail elsewhere[21]. The kicker magnet chain normally kicks the beam into the 10 m long A0 abort absorber[22]. A horizontal bend of 15.7 mrad is required to cross the switchyard beam at PV92 and also to miss the abort dump. A vertical bend of 4.7 mrad is also needed leading to a combined bend of 16.4 mrad. Both bends could be provided by tilting the crystal just as is done with the skew dipole chain for normal fixed target operations. The extraction and transmission efficiency of the extraction crystal would be about 9% based on the Biryukov E853 simulation[3] and other models of the channeling process[23].

A reasonable place to reconnect with the switchyard beam line is at PV92. This would require a horizontal bend of 7.17 mrad coupled with a downward vertical bend of 2.17 mrad. This could be done with several dipole magnets or a second crystal.

Crystal extracted beam intensity is related to the circulating proton beam loss rate. There is a natural loss rate that occurs due to the beam-beam interaction at the collision points, gas scattering, nonlinearities, and other sources. At the Tevatron collider the beam-beam interaction is the dominant source of loss for present and future luminosities. These natural losses must be controlled by scraping the circulating beam. In fact a scraping crystal could cut the loss rate at the detectors. Such a possibility was discussed for the SSC[24].

The external beam rate is then a function of the beam loss in the accelerator, the fraction of the beam losses that strike the crystal, the overall extraction and deflection efficiency of the crystal system, and the maximum allowed loss rate at the collider detectors. For E853 the CDF loss limit at their detector was limited to less than 5 Khz with a loss rate of 3.5 KHz without the crystal and the D0 limit was < 1 KHz with a loss rate quite close to that. Six bunches were colliding, the circulating proton beam was $10^{12}$, the proton lifetime was 75 hours, the luminosity lifetime was 15 hours, and the CDF luminosity was typically $0.5 \cdot 10^{31}$.

A good extracted rate for E853, which kept the D0 loss within its limit was 0.15 MHz. The extraction efficiency was of the order of 25%, so that 0.6 MHz was incident on the crystal. After the Main Injector comes on-line, there will be 36 bunches rather than 6 and the proton intensity will increase tenfold. It is reasonable to expect that the beam incident on a crystal could be 6 MHz. An optimized crystal length of 12.5 cm results in an extraction efficiency of 9%, so that the beam transmitted down the Switchyard line would be 0.5 MHz, or 0.3 protons per bunch per turn.

This design could be improved in several ways. A germanium crystal[25] with its larger critical angle would help. A smaller crystal bending angle could feed a septum magnet chain such as the Lambertson string that will no longer be needed for collider operation. These two improvements could result in a gain of 5 to 10. Another possibility would be to cool the crystal[26]. This would increase the dechanneling length and improve the extraction efficiency, perhaps by a factor of two. Finally a pre-scatterer such as a one-quarter channeling wavelength crystal[27] or an amorphous target[16] might raise the yield by a factor of 2. With these improvements a 5-10 MHz beam is within reach.

There was no evidence of radiation damage in 70 hours of operation for E853. An extraction crystal would have to last about 100 times longer. Measurable damage[28] with an effect on channeling occurs at a rate on the order of $10^{20}$ protons/cm$^2$. Assuming the active region of the crystal is 10 microns wide and 0.6 mm high, an extraction crystal should be able to handle on the order of $10^{16}$ particle passes or years of parasitic operation.

### B. A charm experiment using a crystal extracted beam

The possible application of a crystal extracted beam for a fixed target heavy flavor experiment has been explored in some detail for the SSC, LHC, and the Tevatron[29]. The Fermilab fixed target experiment E771 produced $1.6 \cdot 10^9$ charm pairs during a 30 day run in a 40 MHz beam with a 5% interaction length target. Based on that a 10 MHz parasitic crystal extracted beam could produce $10^{10}$ charm events in a year. Physics possibilities include observations of mixing in the charm sector and CP violation in $D^0$ decays.

### C. Measuring charm baryon magnetic moments

Charm baryon magnetic moment measurements could provide direct information on the magnetic moment of the charm quark. Short charm lifetimes make such measurements difficult with conventional techniques. With the high effective magnetic fields available in a bent crystal charm moment measurements are conceivable[30]. An experiment at Fermilab[31] used this approach to measure the magnetic moment of the $\Sigma^+$. Optimistically a 6% measurement of the $\Lambda_c^+$ magnetic moment would require 3000 $\Lambda_c^+ \rightarrow p + K^- + \pi\pi^+$ events. This would need $4 \cdot 10^{13}$ protons or about a month of running in a 10 MHz beam. One distinct advantage of a parasitic extraction beam for the experiment would be the lower proton rate on target because the beam live time would be 100% rather than the normal 33% for the Tevatron fixed target program.

### XII. SUMMARY

E853 has demonstrated that useful TeV-scale beams can be extracted from a superconducting accelerator in the course of collider operation. An important aspect of this is multi-turn extraction, which substantially increases the efficiency of the process. A significant luminosity-generated halo created by normal collider operation was observed and extracted. This was done without unduly affecting the backgrounds at the collider detectors. The experiment was able to study both luminosity-driven extraction and multi-turn effects in enough detail to confidently apply the technique to real situations.

Extraction efficiency was studied in some detail but proven rather hard to measure accurately during a tightly-packed accelerator schedule where much of the E853 running was parasitic. Nevertheless it is clear that the extraction efficiency is consistent with simulations and that the simulations can be used to extrapolate to future beam designs.

Time dependent effects in the experiment have provided interesting insights into several crystal extraction and accelerator beam halo phenomena.

In kick mode the major time structure was due to accelerator beam dynamics. The turn-to-turn microstructure was sensitive to accelerator tune and more indicative of tune changes than channeling phenomena. Following an initial decay with a time constant on the order of 1 ms the extraction signal was almost flat and then dropped in the 30 ms region. Small, longer term extraction pulses occurred out to 5 seconds. Kick mode extraction was reasonably modeled by the simulation programs except for the long-term persistence and the fact that so-called wrong side turns came into equilibrium with right side turns after a time constant comparable to the initial fall time for right-side buckets. These effects may have been due to non-linearities in the accelerator.

The time for an extracted signal to appear in kick mode for a misaligned crystal was also studied. This possibility arose because of multipasses through the crystal and the associated multiple scattering. The times are consistent with the simulation and also with a simple picture of the process based on multiple scattering.

Other E853 observations related to the character of the circulating accelerator beam halo. Retracting the crystal or a collimator shed light on the diffusion rate of the halo. This rate was related to the beam halo phase space density and non-linear effects in the accelerator system. Typically a 200 μm crystal pullout dropped the extraction rate by a factor of 4 after which it increased with a time constant of several minutes. Based on this information it is difficult to speculate on how halo diffusion depends on the pullout distance.

Related halo information was provided by measurements of the number of kicks required to move the edge of the beam out to the crystal. Typically with a 10 KV kick it took 5 to 7 kicks to grow the beam.

When the crystal was moved in the rate rose instantaneously and then relaxed in less than a minute to an equilibrium point with a much longer decay period. After the initial rapid fall the characteristic rates for luminosity on-off and the proton-only cases typically dropped with a 1.5 hour decay constant.

Using kick mode little evidence was found for oscillations and modulations beyond some indication of a weak modulation with a period in the range of 30-50 ms. This modulation could have been due to any of several effects including sub-harmonics of the line frequency or synchrotron oscillations.

Use of transverse white RF noise from a horizontal damper at F11 increased the extraction rate dramatically. Little interesting was learned about time-dependent effects from the damper studies. Equilibrium was reached in less than ten seconds when the damper was turned on or off.

The impact of an amorphous fiber pre-scatterer on crystal extraction was studied by observing the extracted beam produced by the carbon fibers used for the accelerator flying wire system. The fibers produced significant extracted fluxes consistent with expectations.

This experiment demonstrated luminosity-driven extraction for the first time by simultaneously monitoring colliding and non-colliding proton bunches. The extraction rate was found to be proportional to the luminosity. The luminosity on/luminosity off ratio increased by a factor of 4 for a luminosity of $0.5 \cdot 10^{30}$ cm$^{-2}$s$^{-1}$. This was equivalent to decreasing the goniometer distance from the beam by 0.4 mm or 0.7 $\sigma_x$. The impacts of the luminous regions were similar to the effect of the RF noise source at F11. An RF noise signal of 2.1 V (rms) was equivalent to a luminosity of $0.5 \times 10^{30}$ cm$^{-2}$s$^{-1}$.

One potential application for crystal extraction is the construction of a parasitic collider beam from the Tevatron. A 1000 GeV beam into the fixed target areas would have 25% higher energy than any earlier beam. For an optimized geometry with 6 MHz lost on the crystal a 0.5 MHz beam would be transmitted to a fixed target area. With improvements a 5 to 10 MHz beam would be possible. This beam could be used for heavy flavor experiments such as observations of mixing in the charm sector and CP violation in $D^0$ decays. Another possibility would employ a bent crystal for measurements of charm baryon magnetic moments.

### ACKNOWLEDGMENTS

The authors would like to acknowledge important help received from the Beams, Particle Physics, and Computing Divisions at Fermilab in carrying out this experiment.

*Operated by Universities Research Associates, Inc. under Contract No. DE-AC02-76CH03000 with the United States Department of Energy.

### FIGURE CAPTIONS

FIG. 1. Schematic of the Fermilab accelerator complex showing the location of the crystal extraction experiment at C0.

FIG. 2. Schematic of the channeling extraction experiment at C0. The bent crystal is on the outside of the ring and deflects protons up through the quadrupoles (Q) into the field-free region of the Lambertson magnets. Downstream of the C0 midpoint the extracted protons are detected in two air gaps containing scintillators, a scintillating screen, and a SWIC.

FIG. 3. (a) Orientation of the proton and anti-proton beams at the crystal looking downstream along the proton beam. The crystal deflects up. (b) Beams at the face of Lambertson magnet string.

FIG. 4. Extraction rate in diffusion mode as the crystal was moved in closer to the beam. X is the distance from the beam centerline to the face of the crystal next to the beam.

FIG. 5. Four point crystal bender. An interaction counter below and downstream of the bender is illustrated schematically (not to scale).

FIG. 6. (a). Goniometer installed in the Tevatron. The crystal is mounted at the upstream of the 1 m long barrel. The motion is controlled by two sets of stepped motions at the ends. (b). Schematic of the relative motion of the crystal to the 4" diameter goniometer barrel and 3" Tevatron vacuum apertures when the crystal was retracted or inserted in the beam.

FIG. 7. Fluorescent screen images of the extracted beam as $\Theta_v$ was swept up through a 230 µrad scan, from 130 µrad below the peak to 100 µrad above the peak. The length of the dechanneling tail grows because the beam spot moves up and the Lambertson magnet aperture eclipses less of it. For scale, the longest visible portion of the dechanneling tail is 3 mm.

FIG. 8. Vertical angular alignment scans for extraction in kick mode (upper panel) and diffusion mode (lower panel). Note that the diffusion curve is wider.

FIG. 9. Interaction counter rate with the background subtracted for a $\Theta_h$ scan in diffusion mode. The downstream end of the crystal points out for positive $\Theta_h$. Note that the minimum of the curve is near $\Theta_h = 0$.

FIG. 10. Comparison of the normalized extracted rate for a $\Theta_h$ scan (circles) to the extraction efficiency from the Biryukov simulation in diffusion mode (diamonds).

FIG. 11. Vertical profile of the extracted beam taken with a thin finger counter. Note the dechanneling tail extending below the main peak. The open circles have been multiplied by 20 to show the behavior of the dechanneling tail. The solid line is a Gaussian fit to the data in the peak region.

FIG. 12. Counter rates for $\Theta_v$ scans for (a) the extracted beam and (b) an interaction counter.

FIG. 13. Extraction rate as a function of the turn number from the computer simulation and from the pulse height in a scintillator following a particular kick, renormalized to the simulation result for turn 2. Pulse heights of less than 30 units are indistinguishable from the noise. There are 10% fluctuations in the relative pulse heights from kick to kick. Agreement between the observations and the simulation is good in the early turns, but extraction persists during later turns longer than the simulation predicts.

FIG. 14. Behavior of the extracted beam signal over a 200 ms period after a kick. Note the initial quick decay, a relatively flat portion, a drop over a 20 ms region, and then the reappearance that hints of an oscillation. The rate is normalized to 1 initially.

FIG. 15. Time to come to equilibrium after a kick with the crystal misaligned by 60 µrad. Each data point is a particular turn. Solid circles are the "right-side" turns. The points with Xs were fitted to an exponential. The rise time constant is 170 µs. The rate is normalized to 1 at the asymptote of the fit.

FIG. 16. Wrong side behavior. Extraction rates on each turn are plotted with the high "right side" pulses circled and the low "wrong-side" pulses marked with squares. The Xs without circles or squares are turns that are in-between "right side" and "wrong side" turns. The fitted curve is discussed in the text. The y axis shows the unnormalized scintillator signal.

FIG. 17. Long term behavior of the extracted beam after a kick. The y axis shows the number of pulses as a function of time after a kick. Crosses give the total number of pulses and circles give the number of low-intensity pulses. Signal intensity decreases with time so that later pulses are mostly small.

FIG. 18. Extraction rate as a function of time immediately after retracting the crystal by 200 μm. The data are fitted with a form $A-Be^{-\lambda t}$ with $t_0 = 1/\lambda = 2.2$ minutes. The filled circles indicate points used for the fit.

FIG. 19. Extraction rate as a function of how far the three collimators had been withdrawn from their normal positions during a major portion of the run.

FIG. 20. Extraction rate as a function of the kick number. Characteristically it took 5 to 7 kicks to reach the equilibrium rate. The extraction rate is normalized to the full extraction rate after the kicks.

FIG. 21. Number of pulses long after a kick as a function of the time within the Main Ring cycle (summed for several kicks). There is little indication of any significant effect of the Main Ring cycle.

FIG. 22. Finger counter coincidence rate as a function of RF damper voltage. The errors are based on inter-point variation. The curve is a fit to the data.

FIG. 23. Smoothed signals for a typical wire fly for extraction (solid circles) and interactions (open diamonds). The extraction rate has been divided by ten. For reference the extracted beam signal for kick mode from Fig. 14 is reproduced here normalized to the flying wire extracted rate (solid line).

FIG. 24. Smoothed time distributions for a wire fly for the prompt signals from extraction (solid circles) and interactions (open diamonds). The extraction rate has been divided by ten.

FIG. 25. Smoothed time distribution for the flying wire interaction rate showing three different regions of exponential decay.

FIG. 26. Illustration of the effect of colliding bunches on count rates for set 1 (CDF and D0 collisions combined) and set 2 (see text). Histograms are measured rates while dots are calculated based on bunch luminosities and a smoothed background. The small open circles on the baselines indicate the times of unmeasured bunches. The vertical axis for set 2 is halved because only one bunch was counted. Numbers above a histogram indicate the proton bunch or bunches.

FIG. 27. Measured, background-subtracted extraction rates normalized to t = 0 versus luminosity. The line is a fit to the data.

FIG. 28. Horizontal and vertical beam layout for a crystal extraction beam at A0. Note the absorber near A0, clearances near XSEM, H90, and V91 at about 150 ft after A0, and the location of PSEP. The scale for x:y is 10:1.

## REFERENCES

[1] S. E. Anassontzis *et al.*, Nucl. Phys. (Proc. Suppl.) **27**, 352 (1992).

[2] S. A. Bogacz, D. B. Cline, and S. Ramachandran, Nucl. Instrum. Methods Phys. Res., Sect. B **111**, 244 (1996).

[3] V. Biryukov, Phys. Rev. E **52**, 6818 (1995).

[4] C. T. Murphy *et al.*, Nucl. Instrum. Methods Phys. Res., Sect. B **119**, 231 (1996).

[5] S. Ramachandran, Ph.D. thesis, UCLA, 1997.

[6] R. Carrigan *et al.*, Phys. Rev. ST Accel. Beams AB **1**, 022801 (1998).

[7] M. Harrison, Fermilab, UPC-153 (1981).

[8] V. Baublis *et al*, Nucl. Instrum. Methods Phys. Res., Sect. B **119**, 308 (1996).

[9] Z. Tang, Fermilab, TM-1827 (1993).

10. X. Altuna *et al.*, Phys. Lett. B **357**, 671 (1995).

11. K. Elsener *et al.*, Nucl. Instrum. Methods Phys. Res., Sect. B **119**, 215 (1996).

12. K. Elsener, W. Herr, and J. Klem, p. 159, **International Symposium on Near Beam Physics**, R. Carrigan and N. Mokhov, eds., Fermilab (1997).

13. A. G. Afonin, et al., Phys. Lett. B **435**, 240 (1998).

14. V. Biryukov, Nucl. Instrum. Methods Phys. Res., Sect. B **117**, 463 (1996).

15. J. Marriner, private communication.

16. A. A. Asseev, et al., Nucl. Instrum. Methods **A324**, 31 (1993) and Nucl. Instrum. Methods **A334**, 283 (1993).

17. V. Biryukov, Nucl. Instrum. Methods **B53**, 202 (1991).

18. A. Taratin, S. Vorobiev, M. Bavizhev, and I. Yazynin, Nucl. Instrum. Methods **B58**, 103 (1991).

19. H. Akbari, et al., Phys. Lett. **B313**, 491 (1993).

20. Richard A. Carrigan, Jr., Nucl. Instrum. Methods **B119**, 239 (1996).

21. R. A. Carrigan, Jr., Fermilab TM-1978 (1996).

22. "The A0 Abort System for the Tevatron Upgrade", C. Crawford, Fermilab TM-1564 (1989).

23. R. Carrigan, p. 339 in **Relativistic Channeling**, eds. R. Carrigan and J. Ellison, Plenum (1987), New York.

24. M. Maslov, N. Mokhov, and I. Yazinin, SSCL-484 (1991).

25. C. Biino, et al., Phys Lett **B403**, 163 (1997).

26. J. S. Forster, H. Hatton, R. J. Toone, G. Este, S. I. Baker, R. A. Carrigan, Jr., W. M. Gibson, R. L. Wijayawardana, J. A. Ellison, L. Emma-Wori, B. O. Kolbesen, Nuclear Physics **B318**, 301 (1989).

27. E. Tsyganov and A. Taratin, Nucl. Instrum. Methods **A363**, 511 (1995).

28. S. I. Baker, R. A. Carrigan, Jr., V. R. Cupps II, J. S. Forster, W. M. Gibson, and C. R. Sun, Nucl. Instrum. Methods **B90**, 119 (1994).

29. D. Christian in "Heavy Quarks at Fixed Target", B. Cox, ed. (Frascati Physics Series), 421 (1994).

30. V. M. Samsonov, Nucl. Instrum. Methods **B119**, 271 (1996). V. V. Baublis, V. G. Riabov, V. M. Samsonov, A. V. Khanzadeev, Preprint PNPI NP-1997 2195 (1997), Gatchina (in Russian).

31. D. Chen et al., Phys. Rev. Lett. **69**, 3286 (1992).

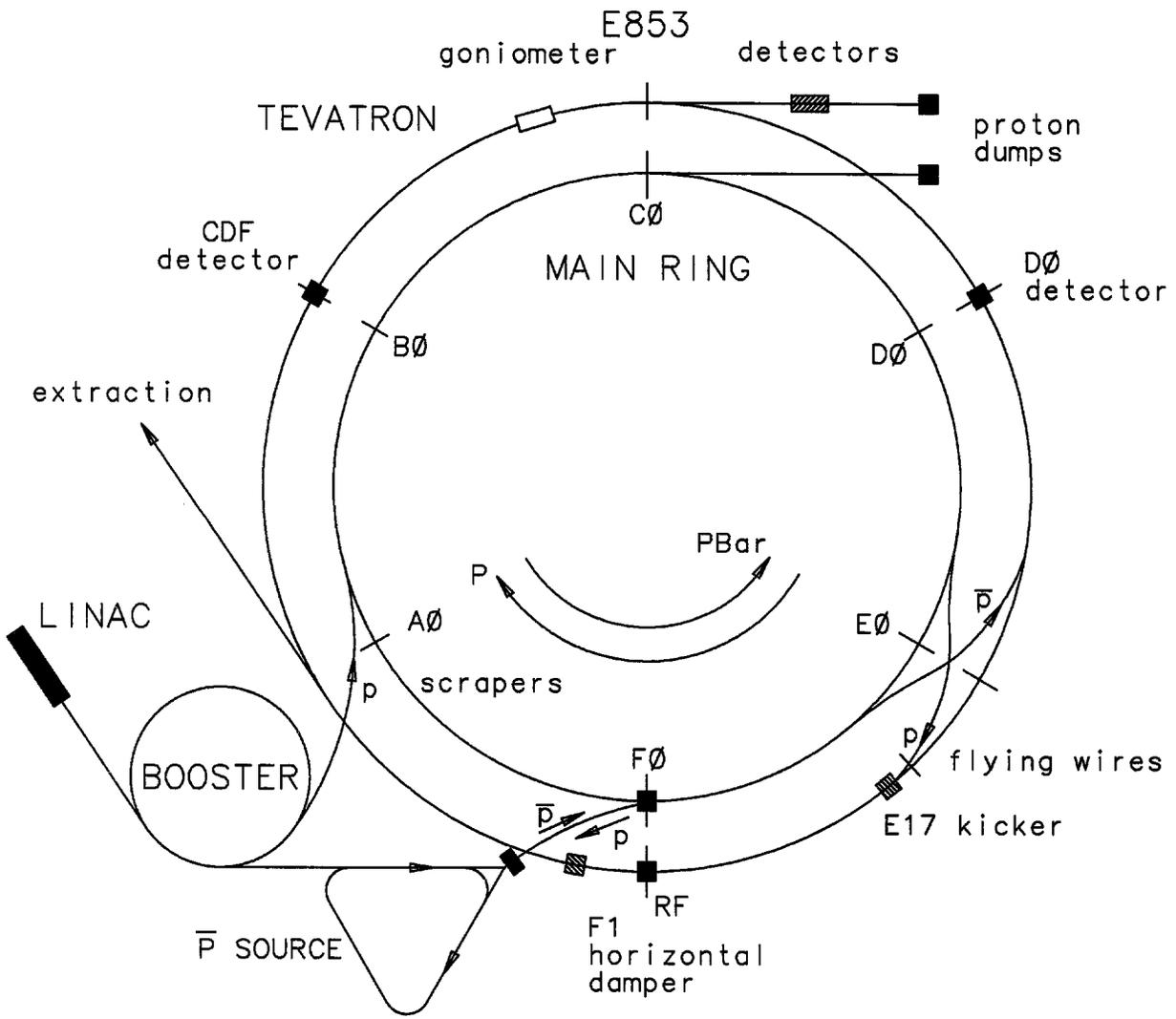

**FIG. 1.**

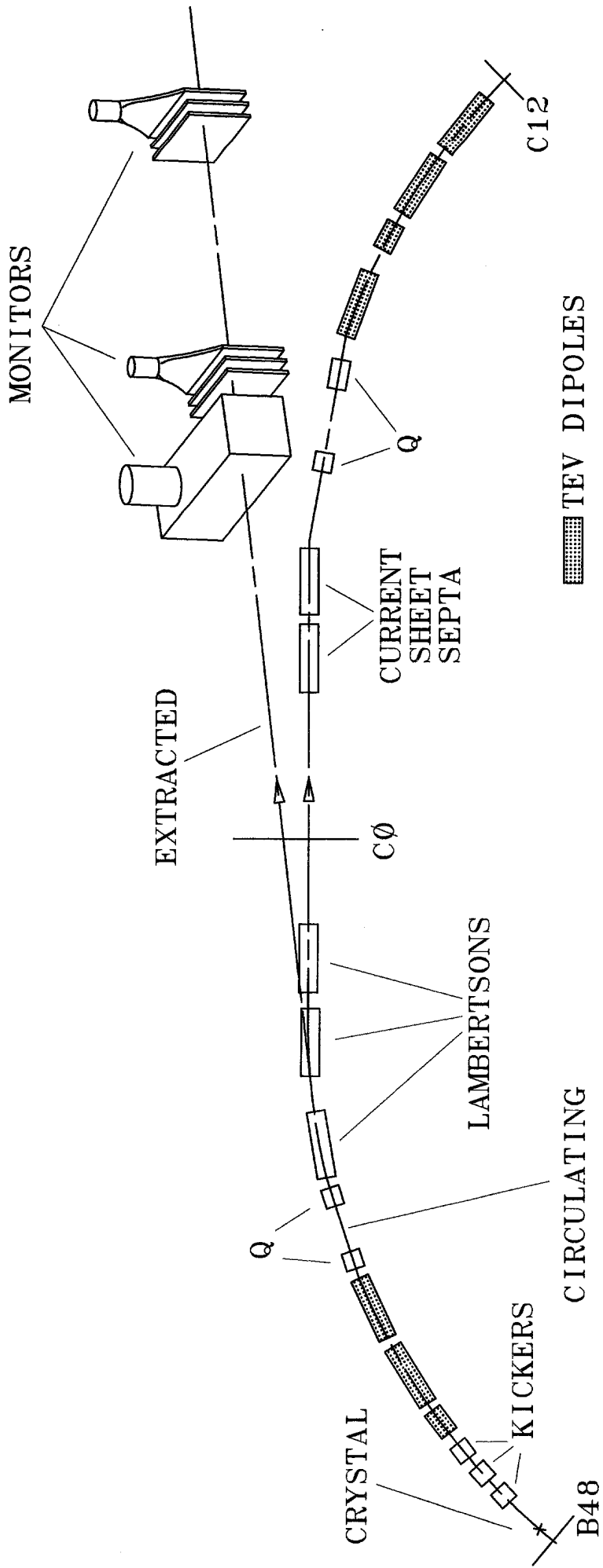

FIG. 2.

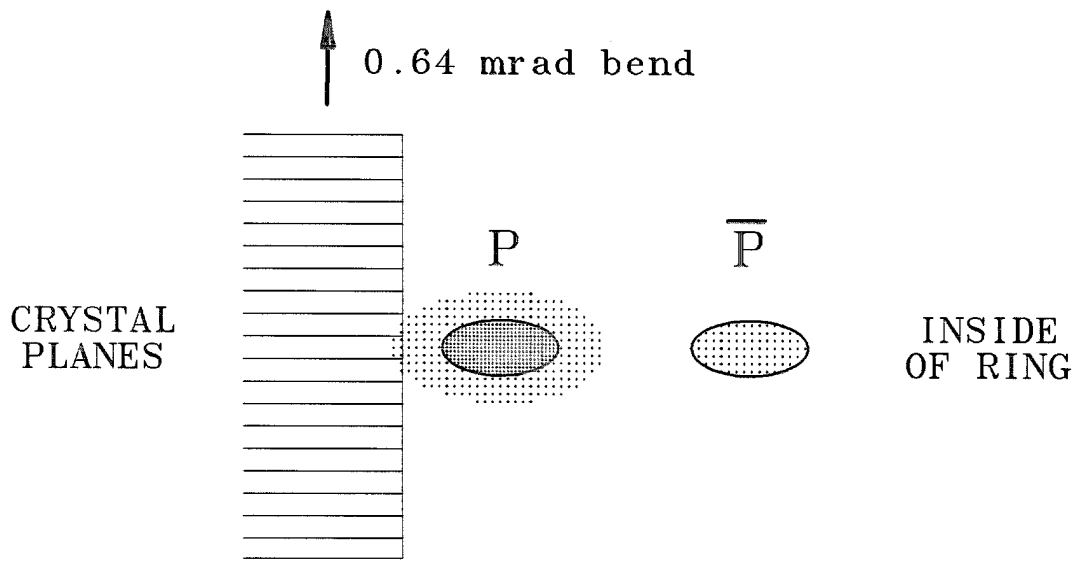

a. detail at crystal

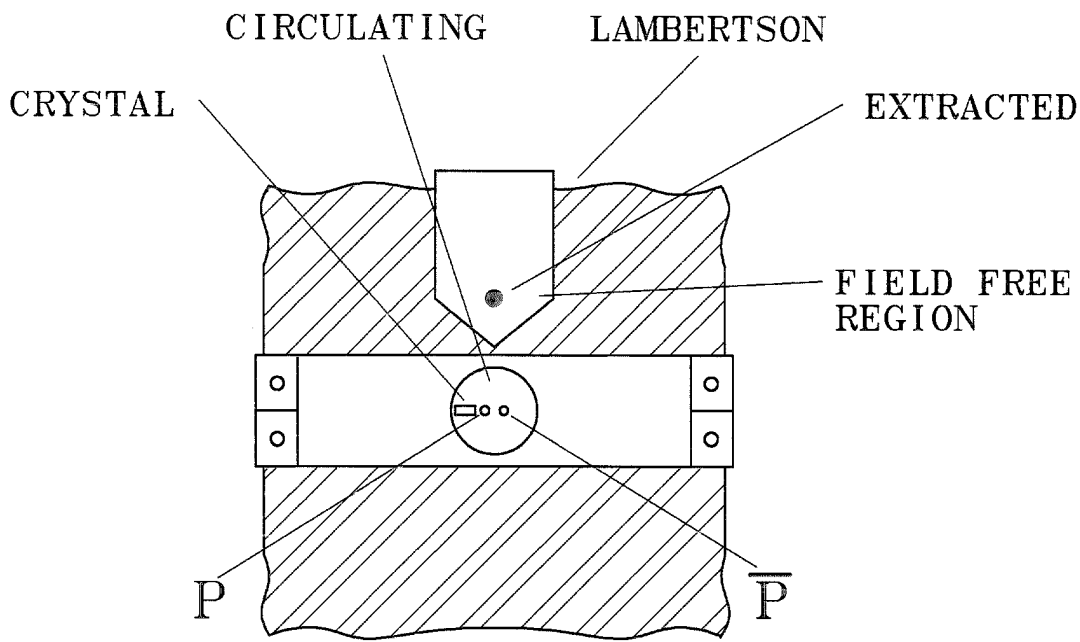

b. crystal and Lambertson

**FIG. 3.**

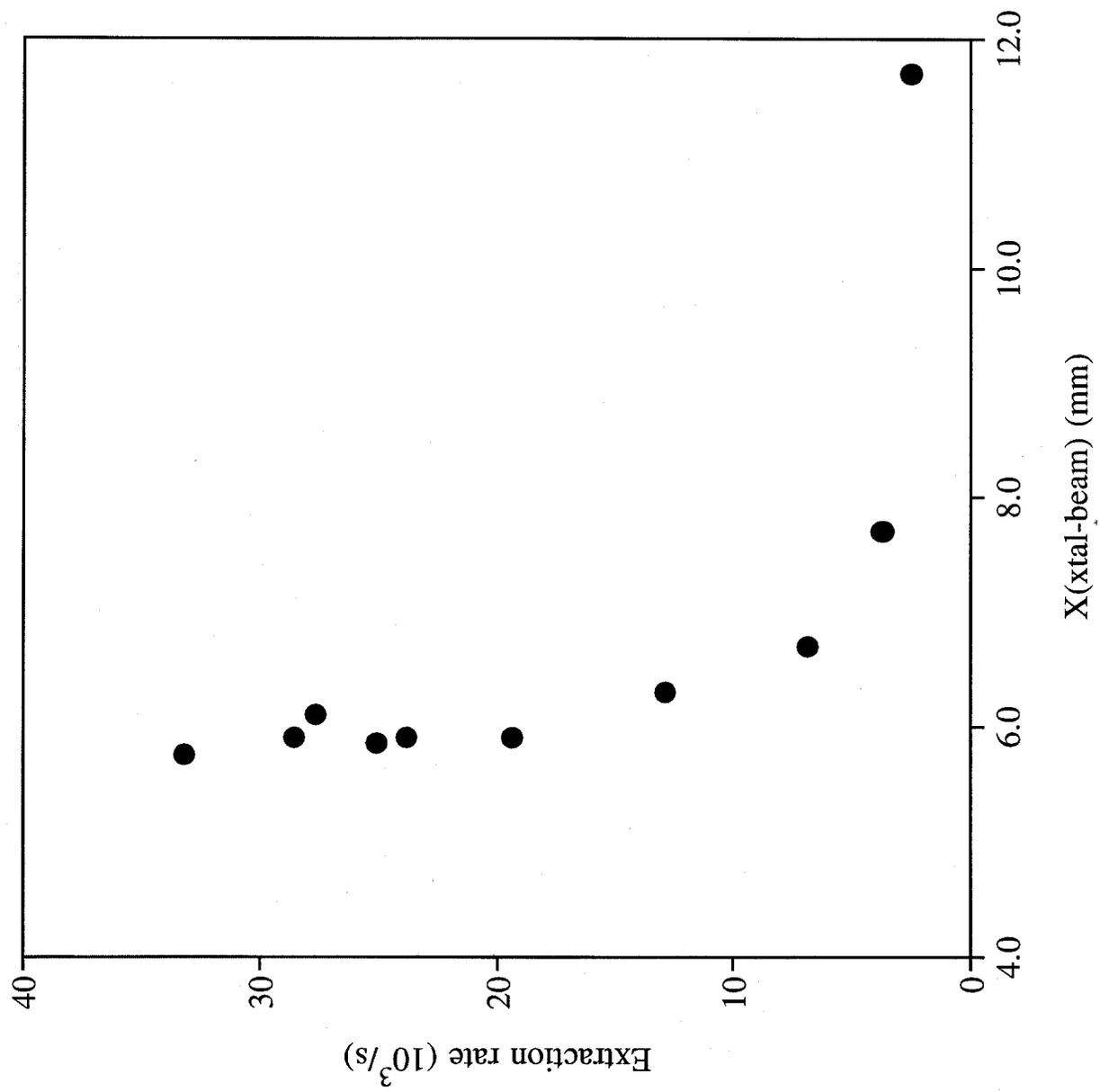

FIG. 4.

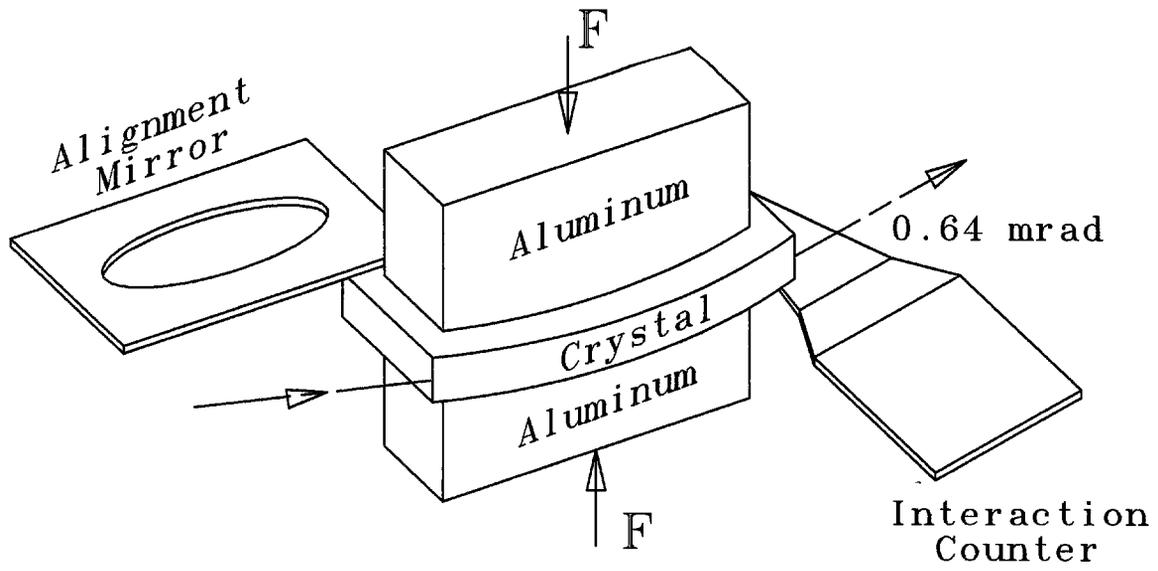

**FIG. 5.**

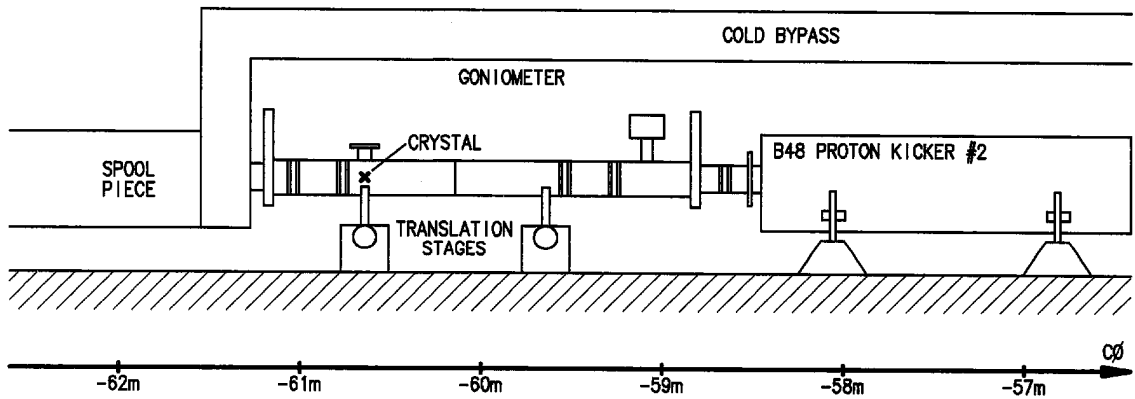

a. goniometer in Tevatron

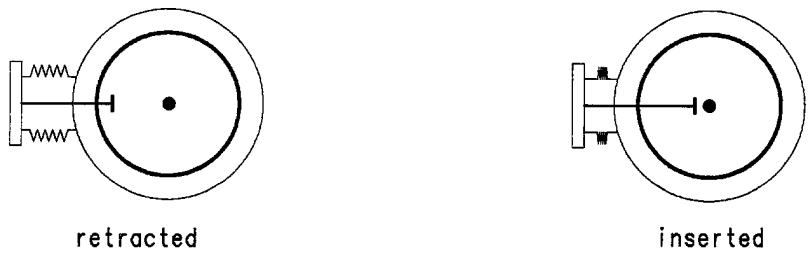

b. position of Tevatron and goniometer

**FIG. 6.**

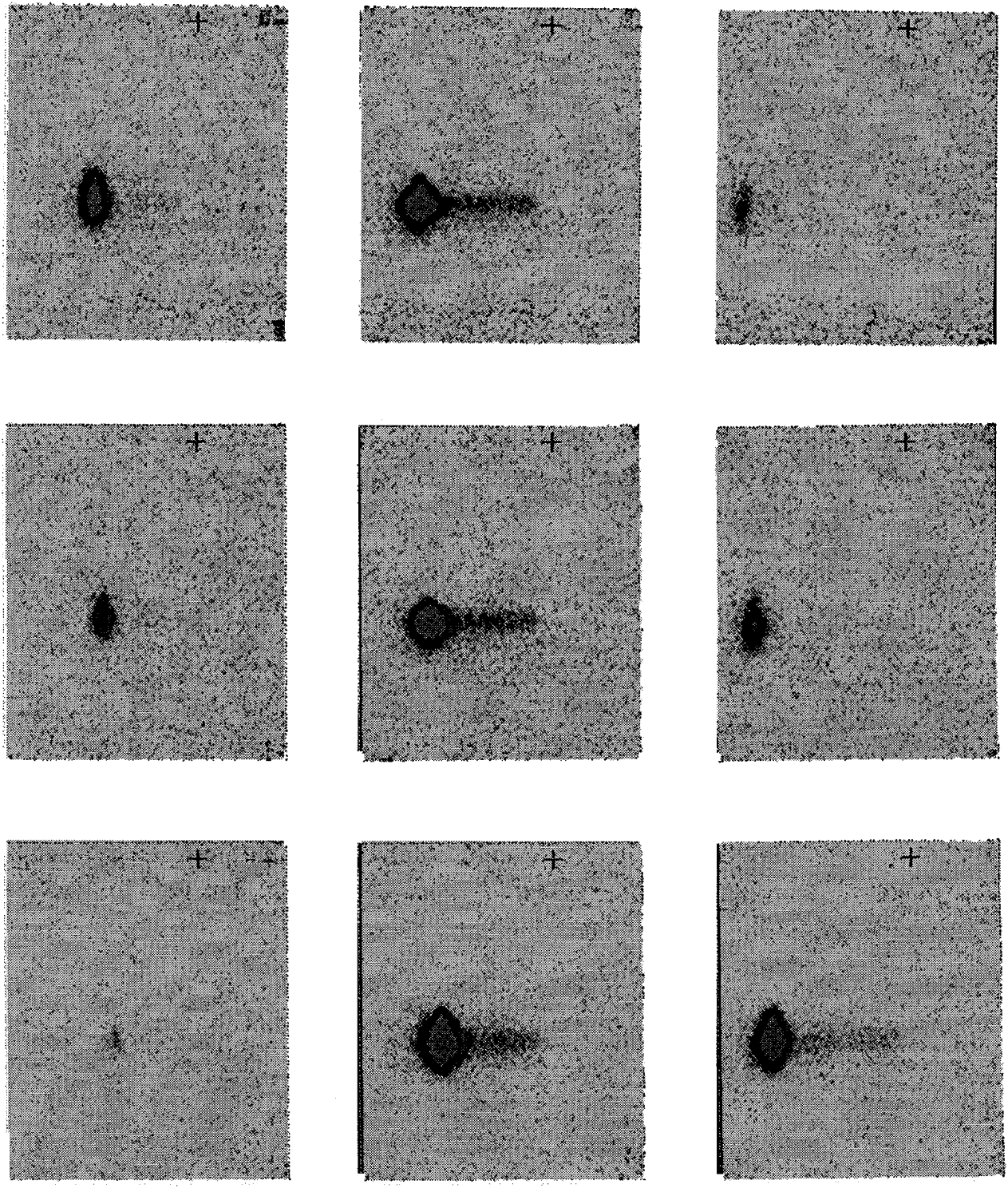

**FIG. 7.**

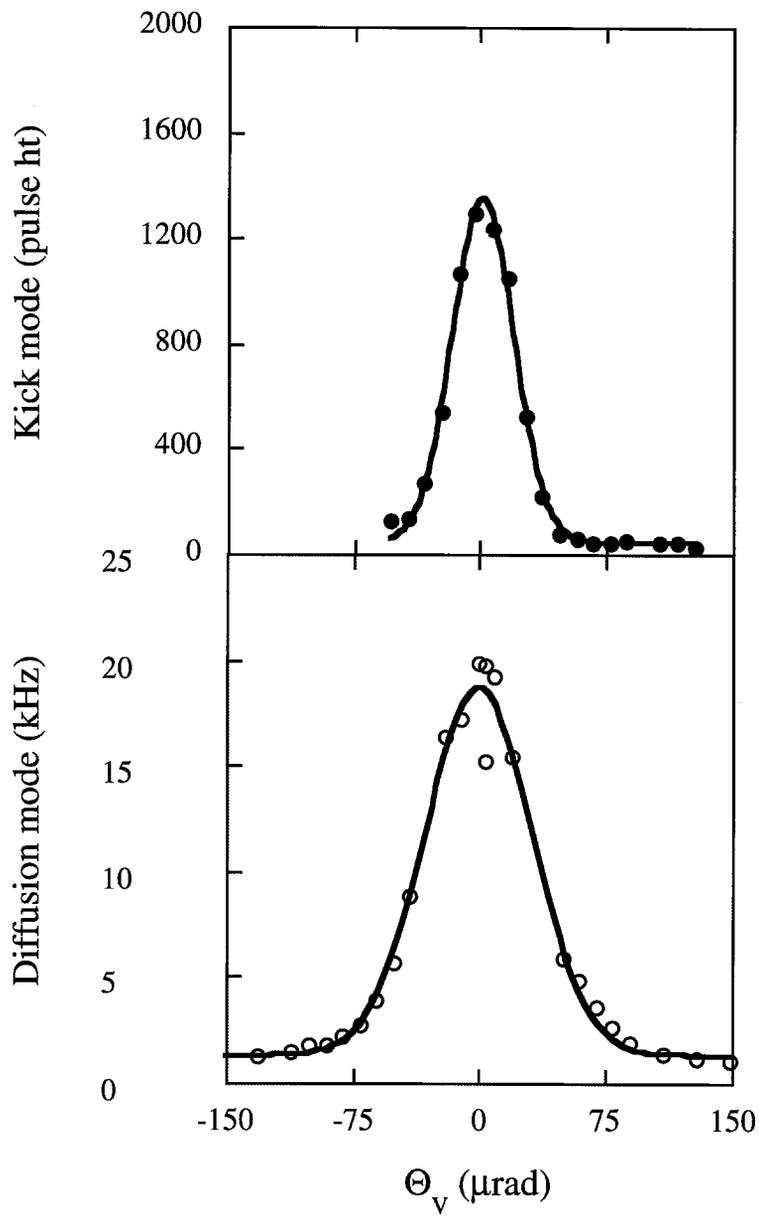

FIG. 8.

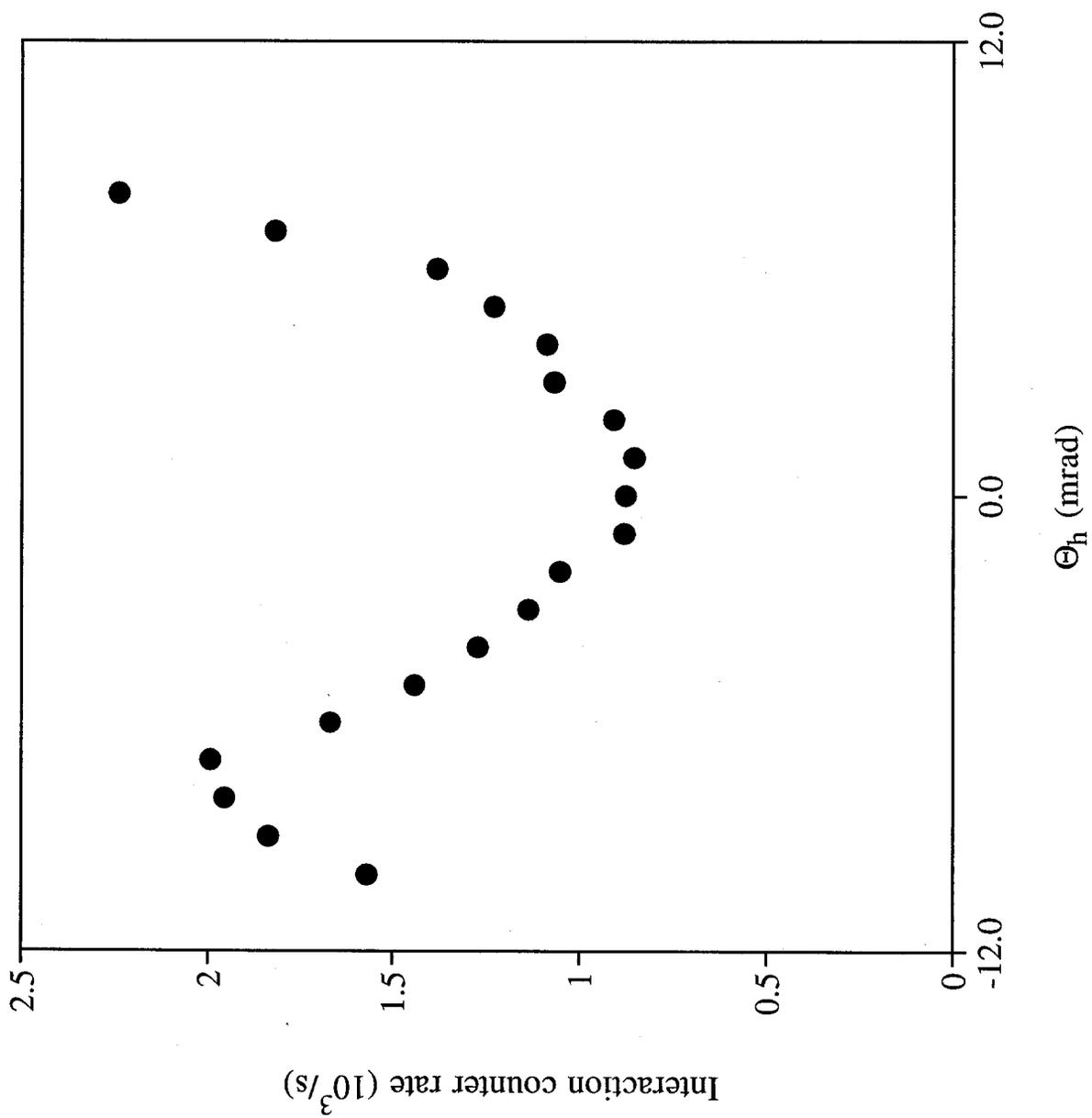

FIG. 9.

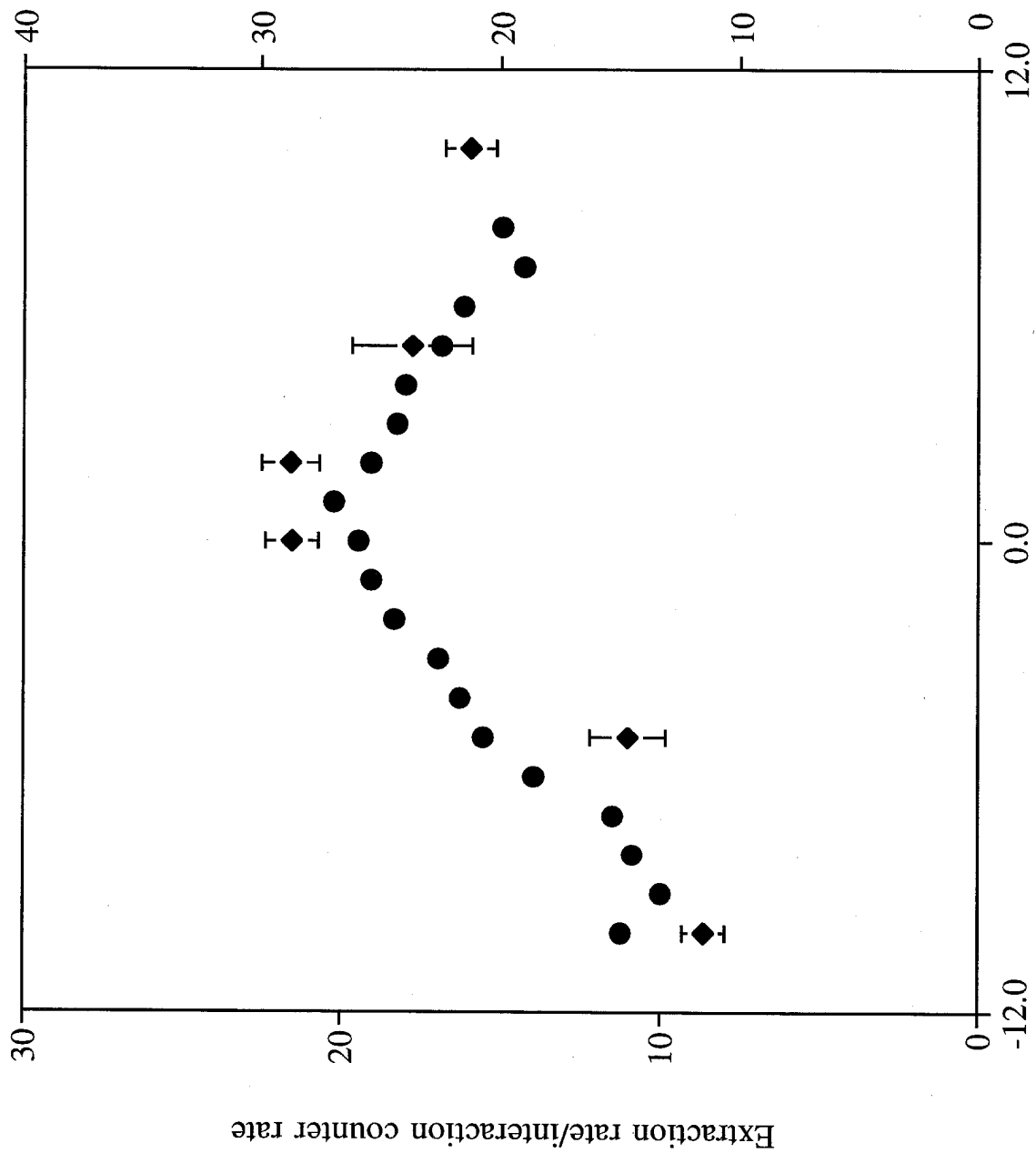

FIG. 10.

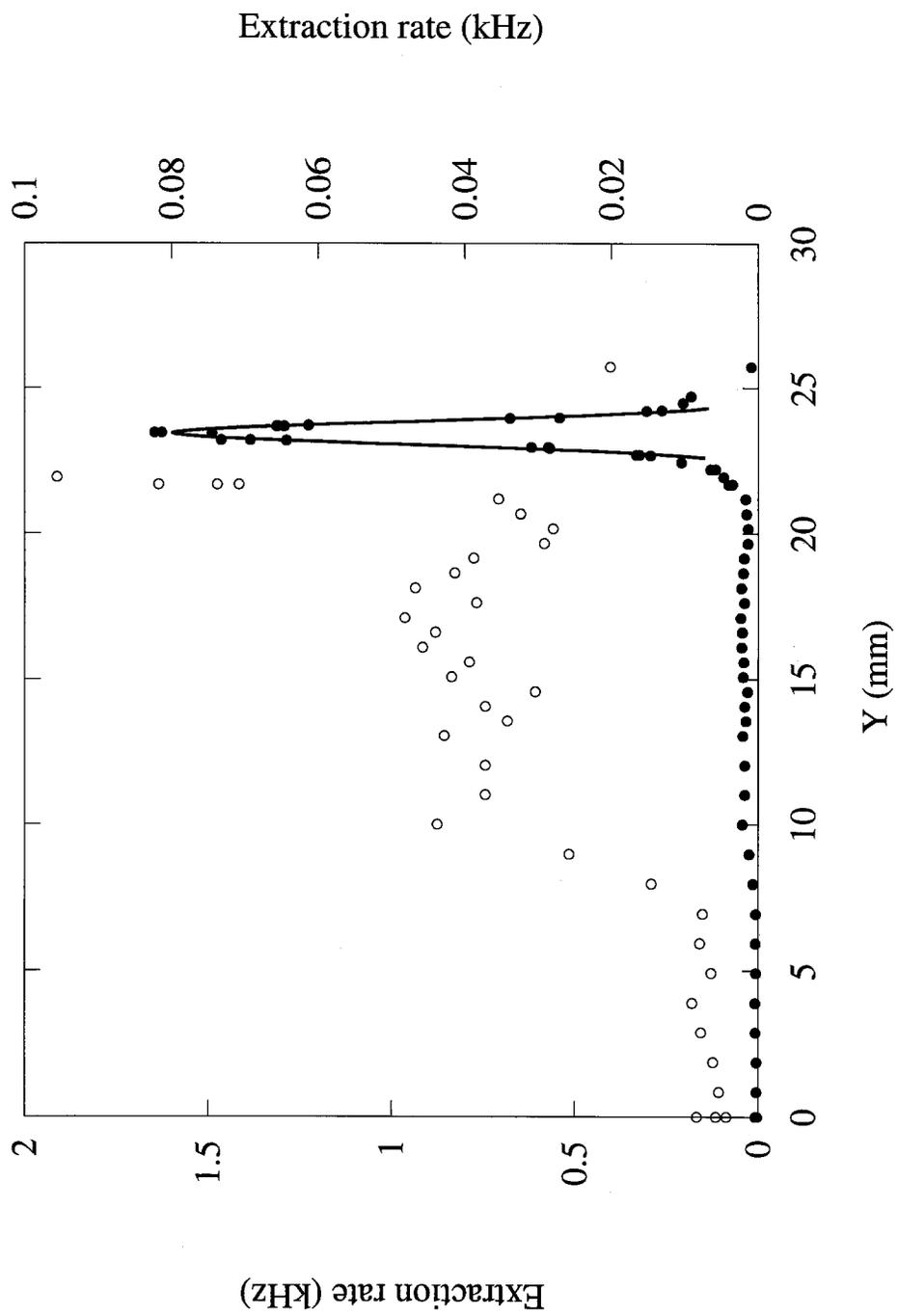

FIG. 11.

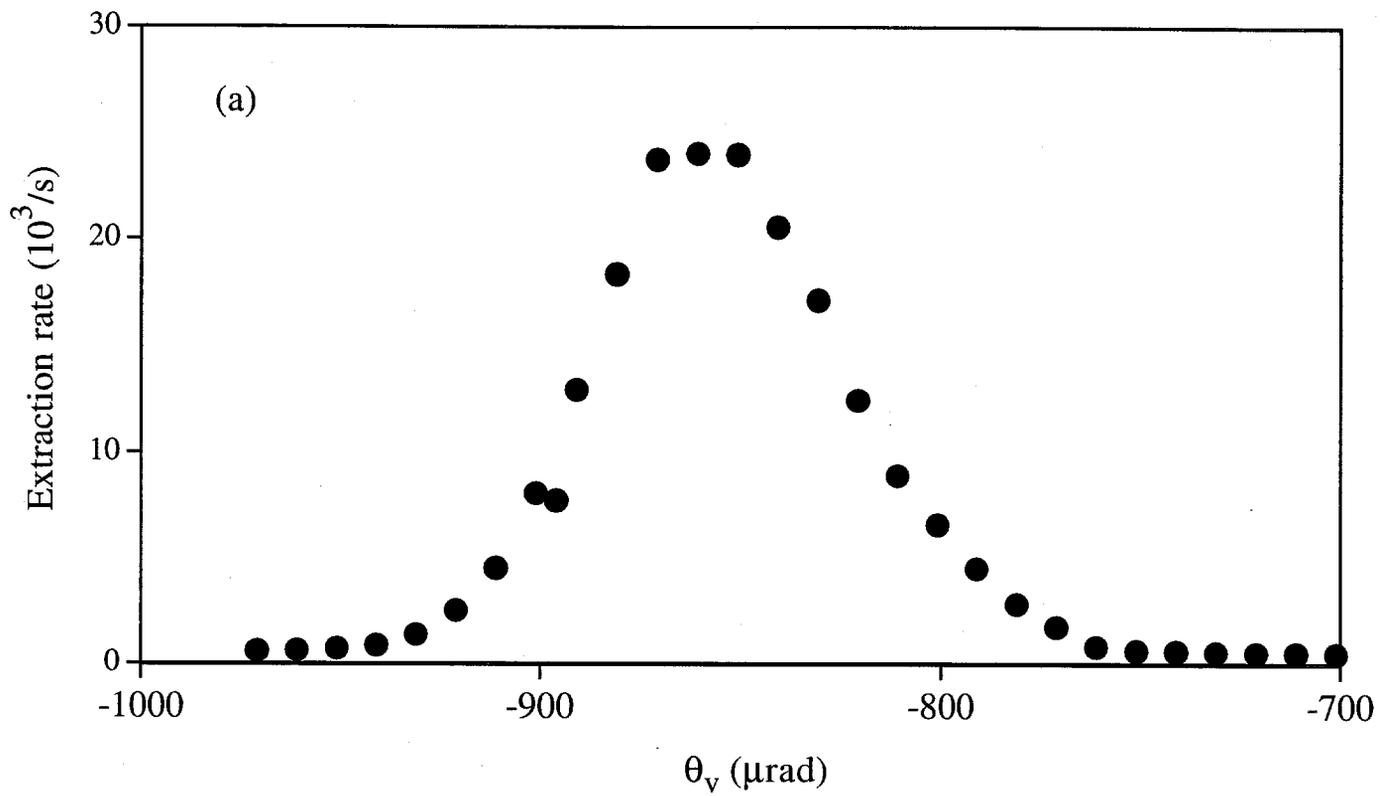
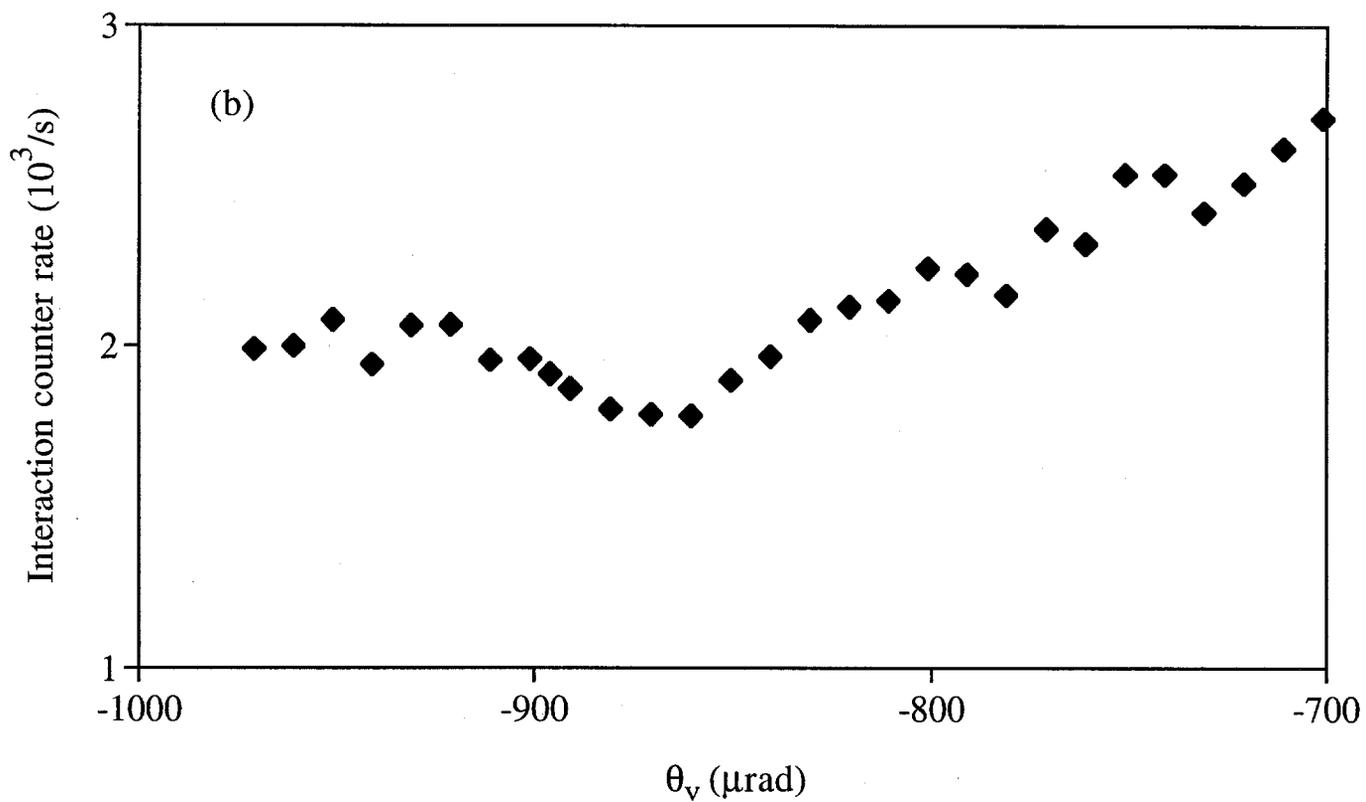

FIG. 12.

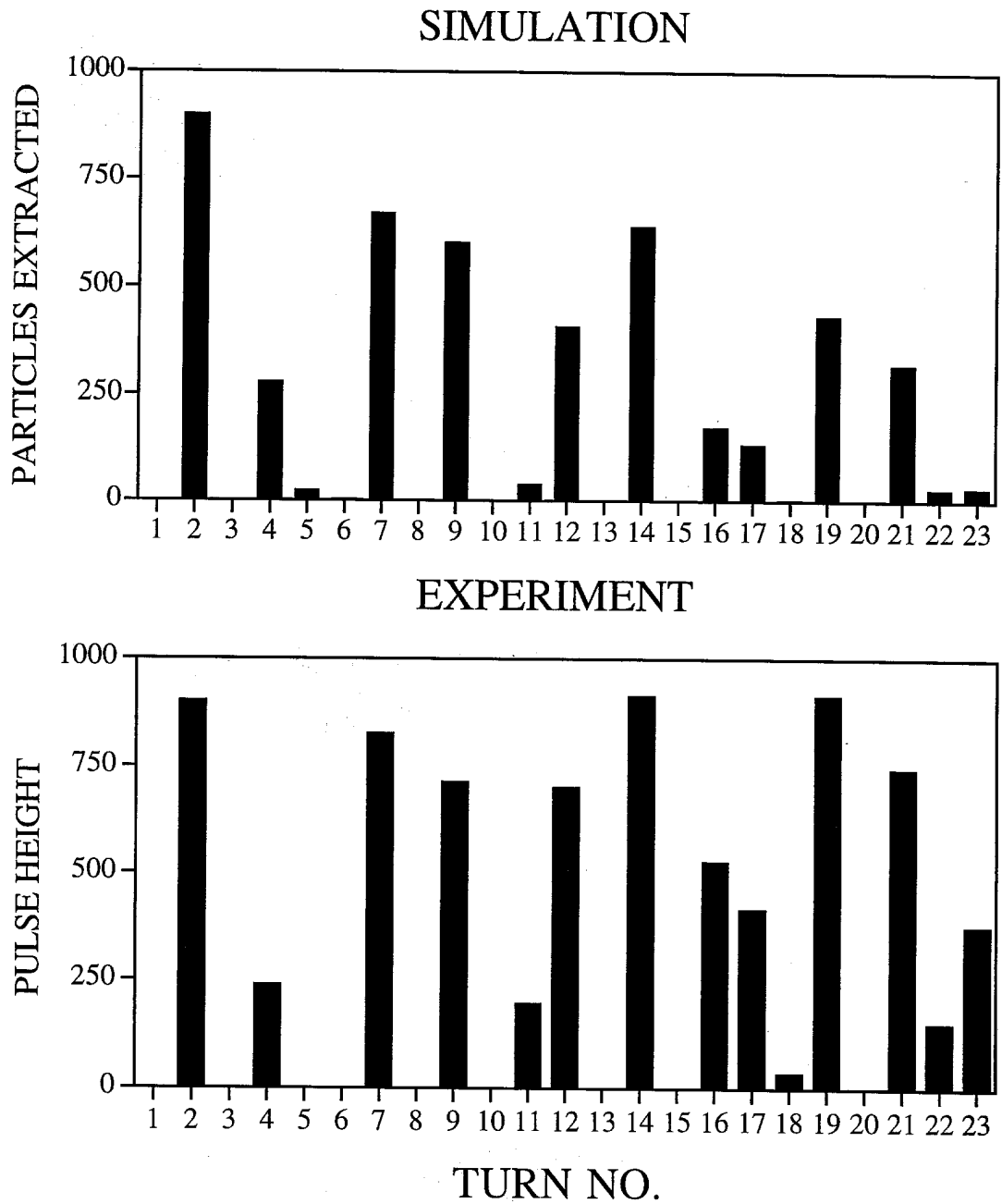

FIG. 13.

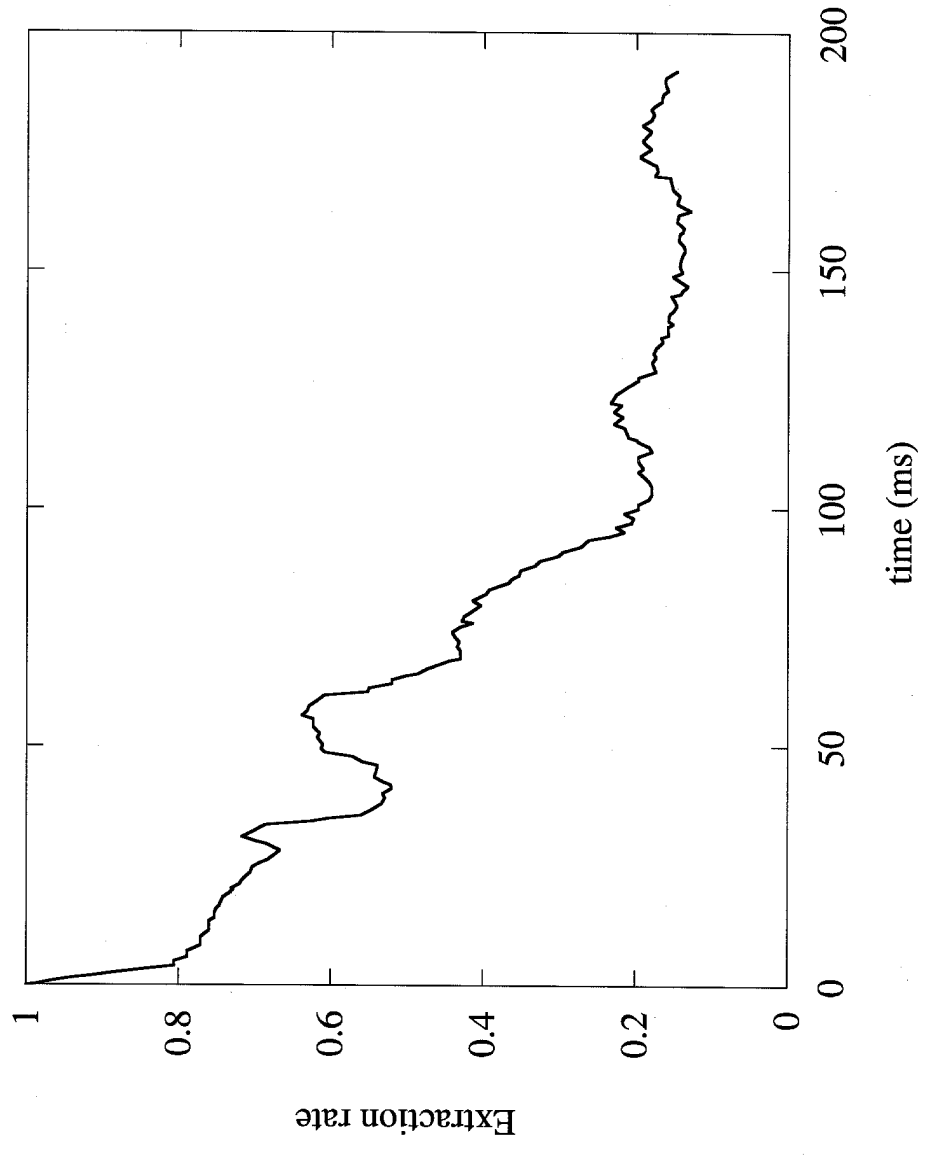

FIG. 14.

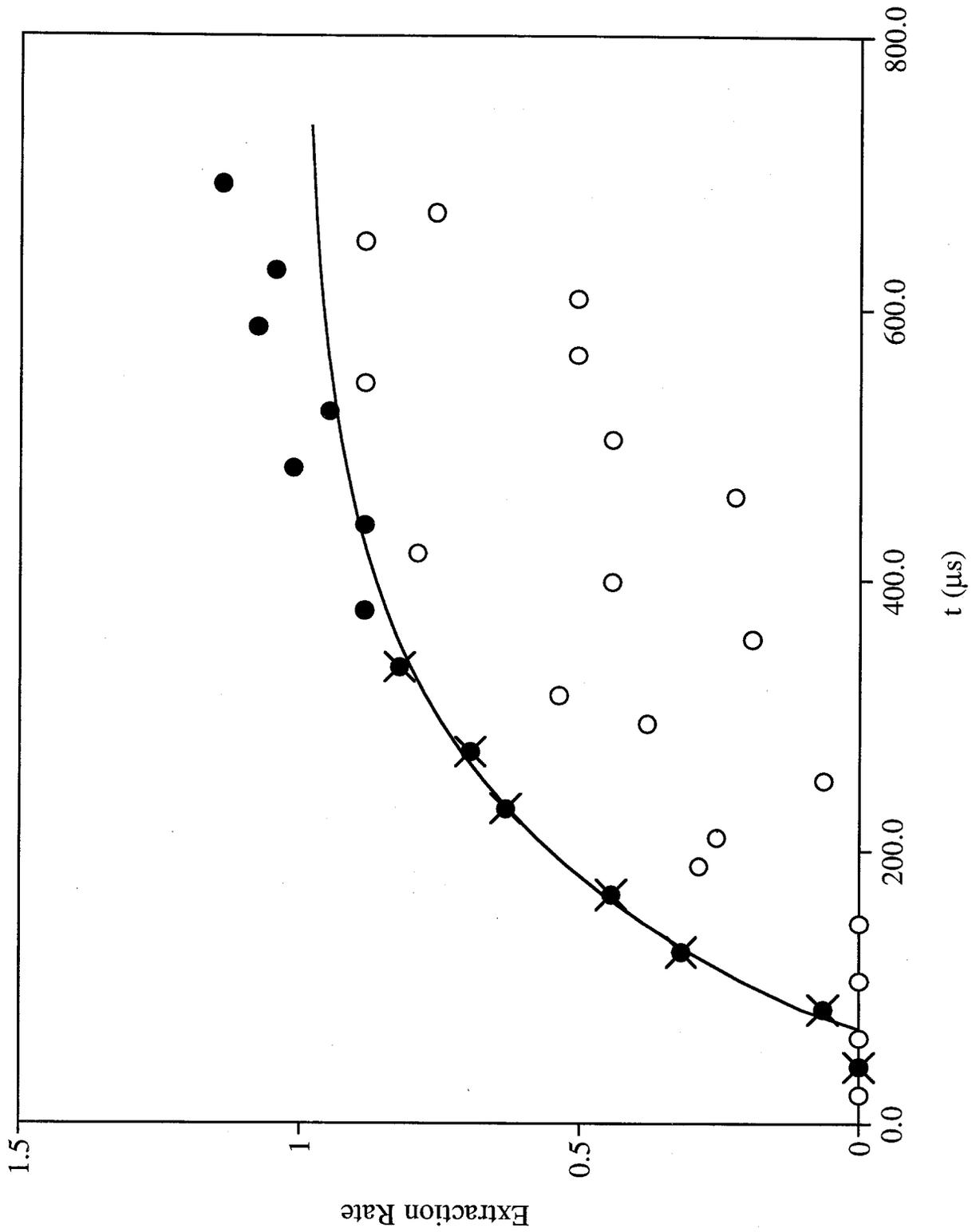

FIG. 15.

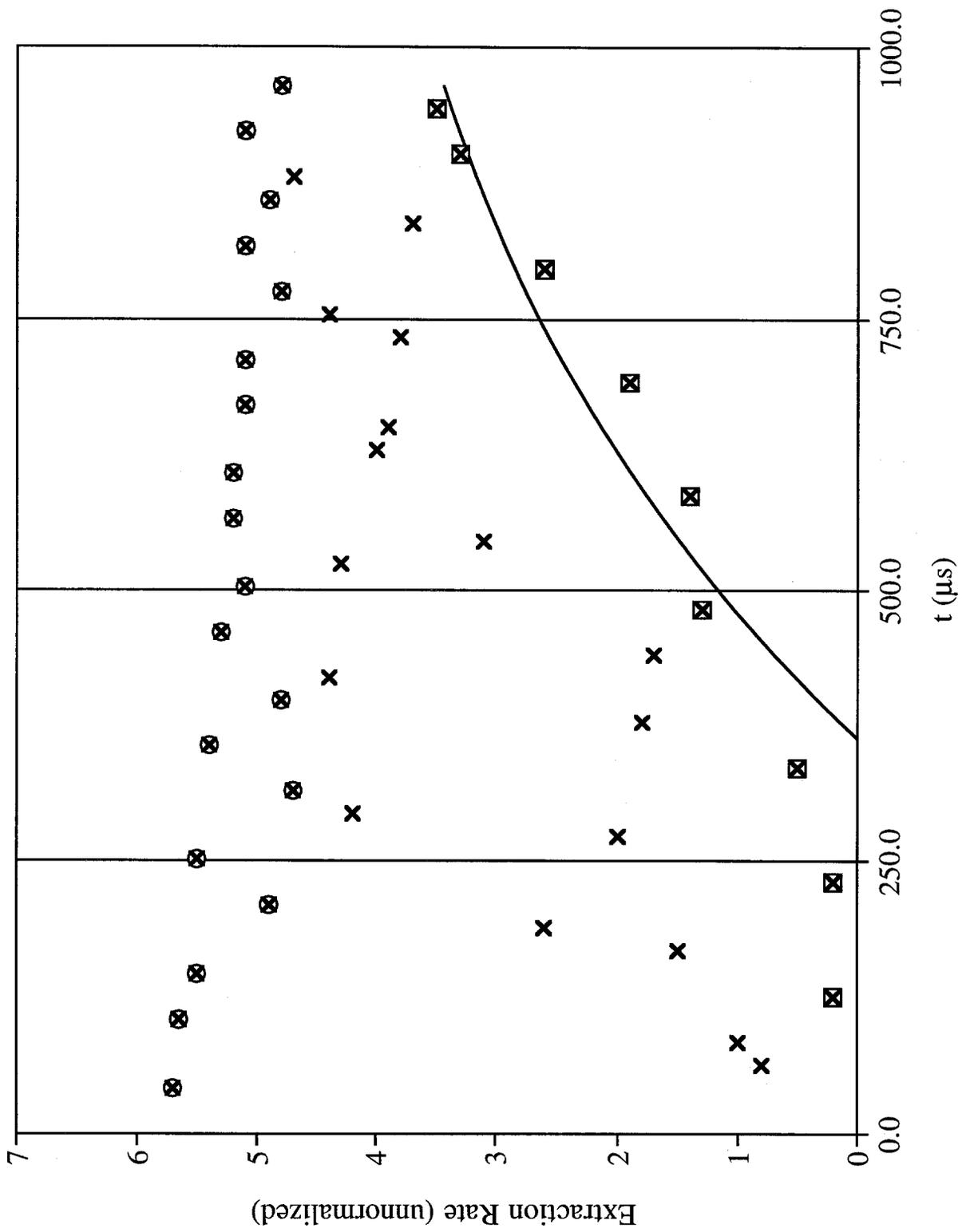

FIG. 16.

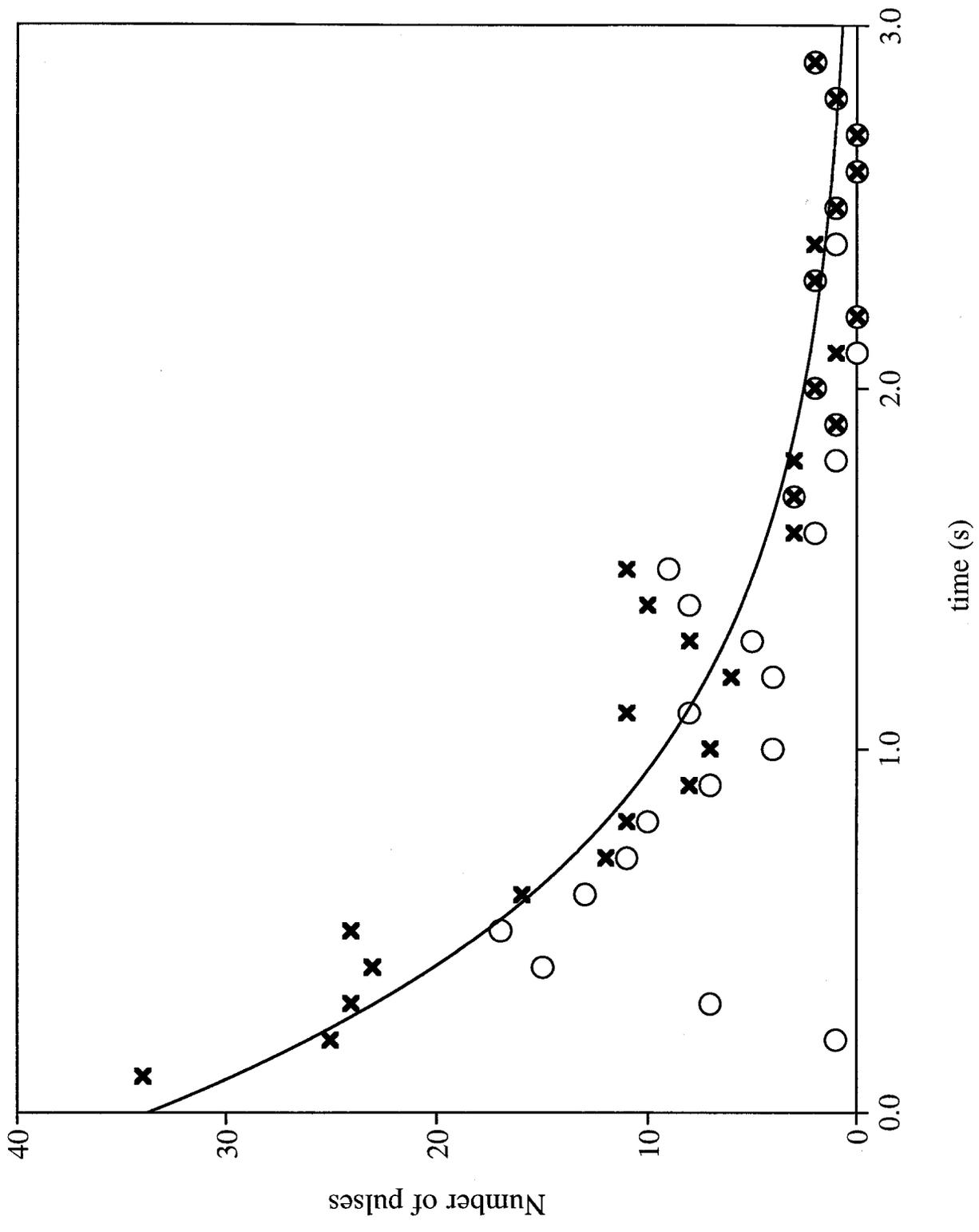

FIG. 17.

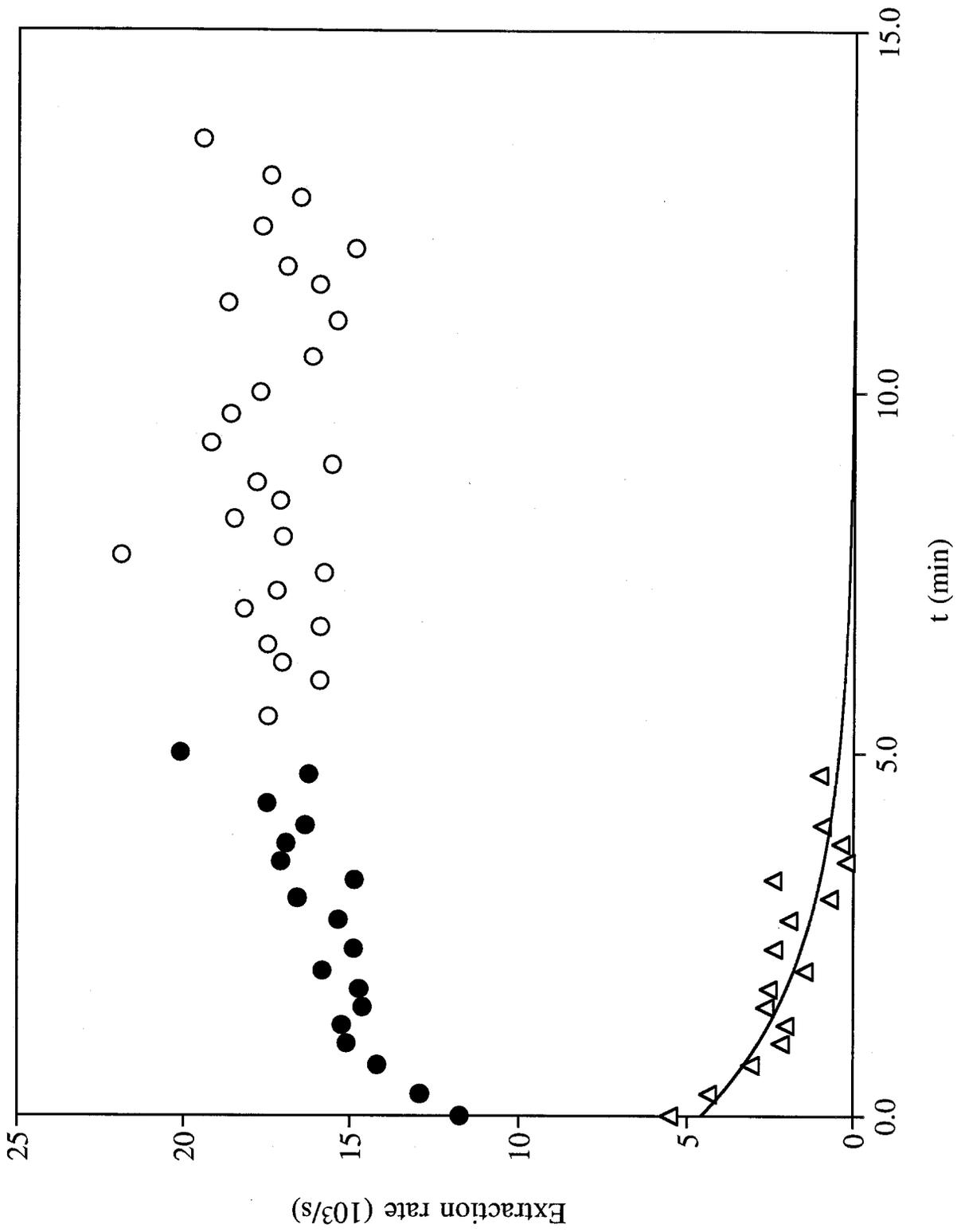

FIG. 18.

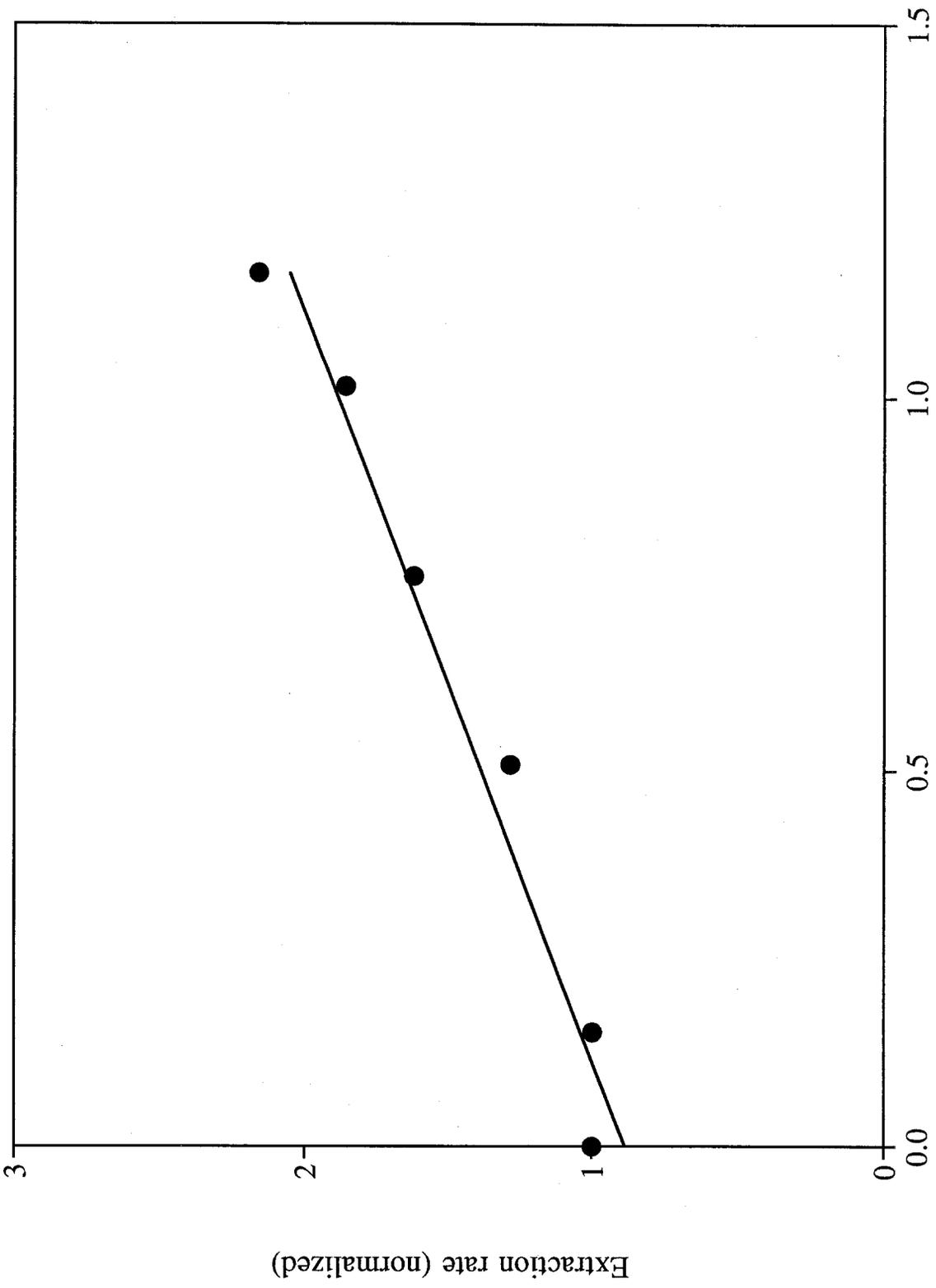

FIG. 19.

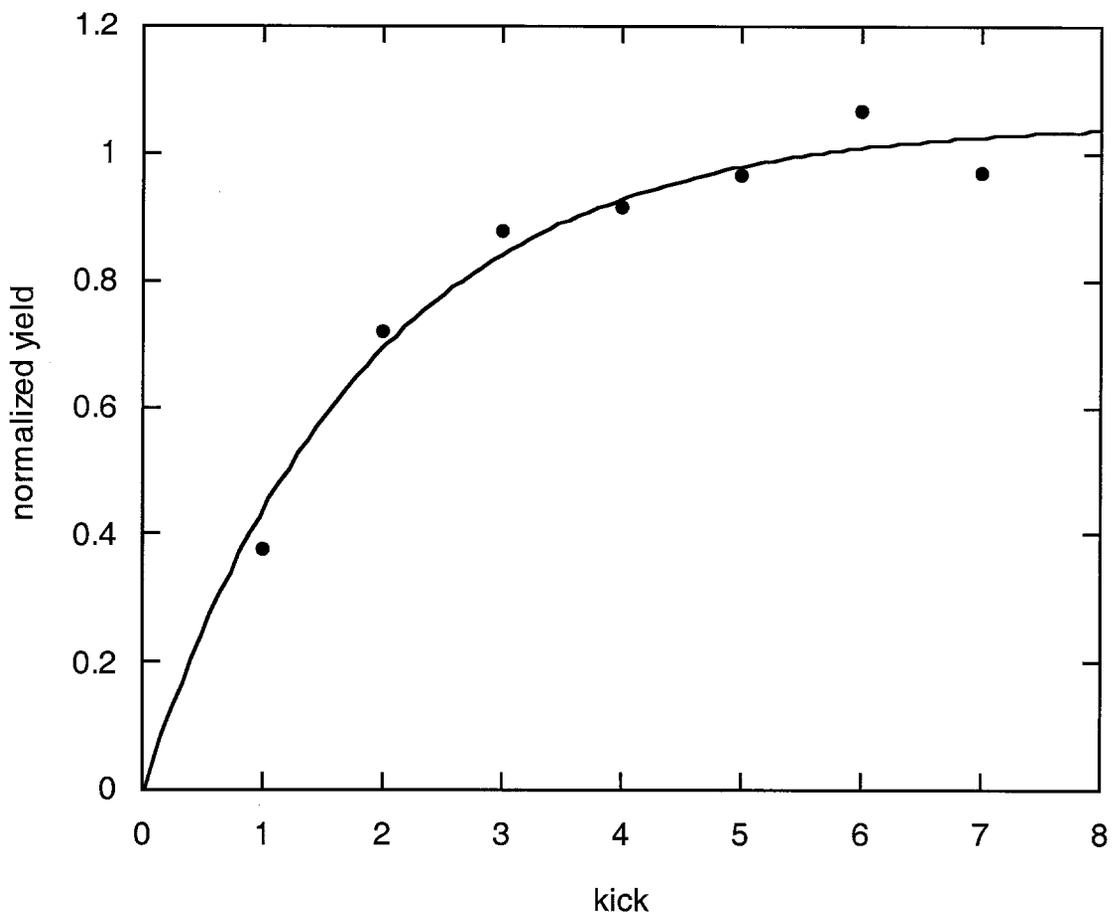

**FIG. 20.**

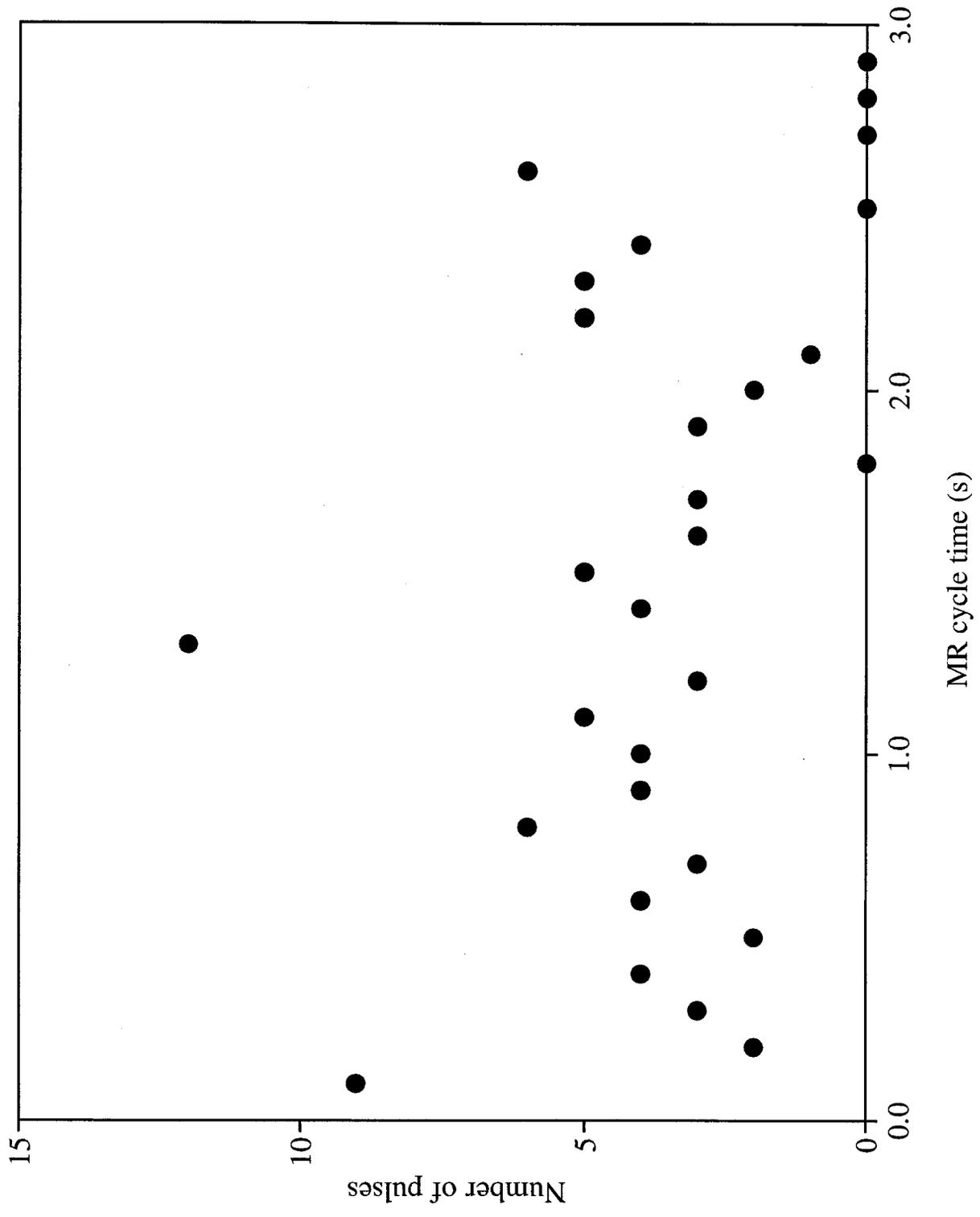

FIG. 21.

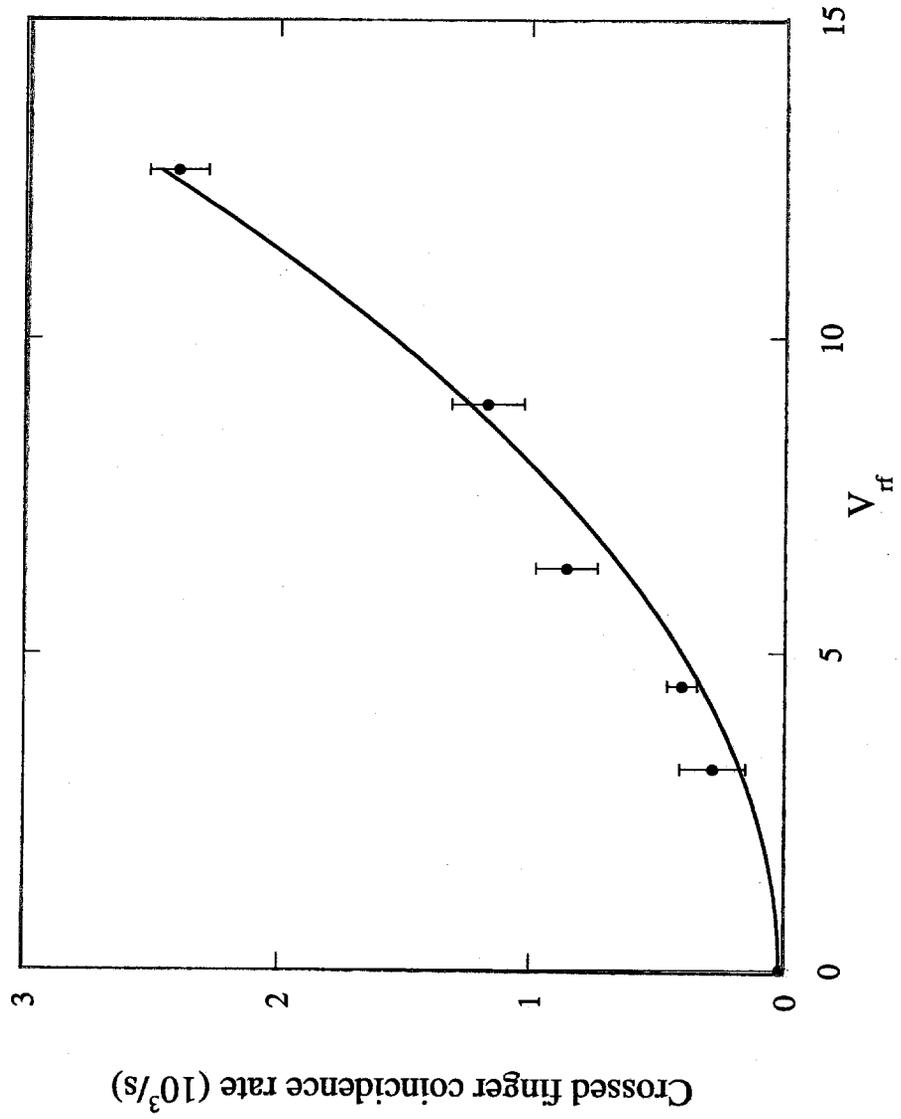

FIG. 22.

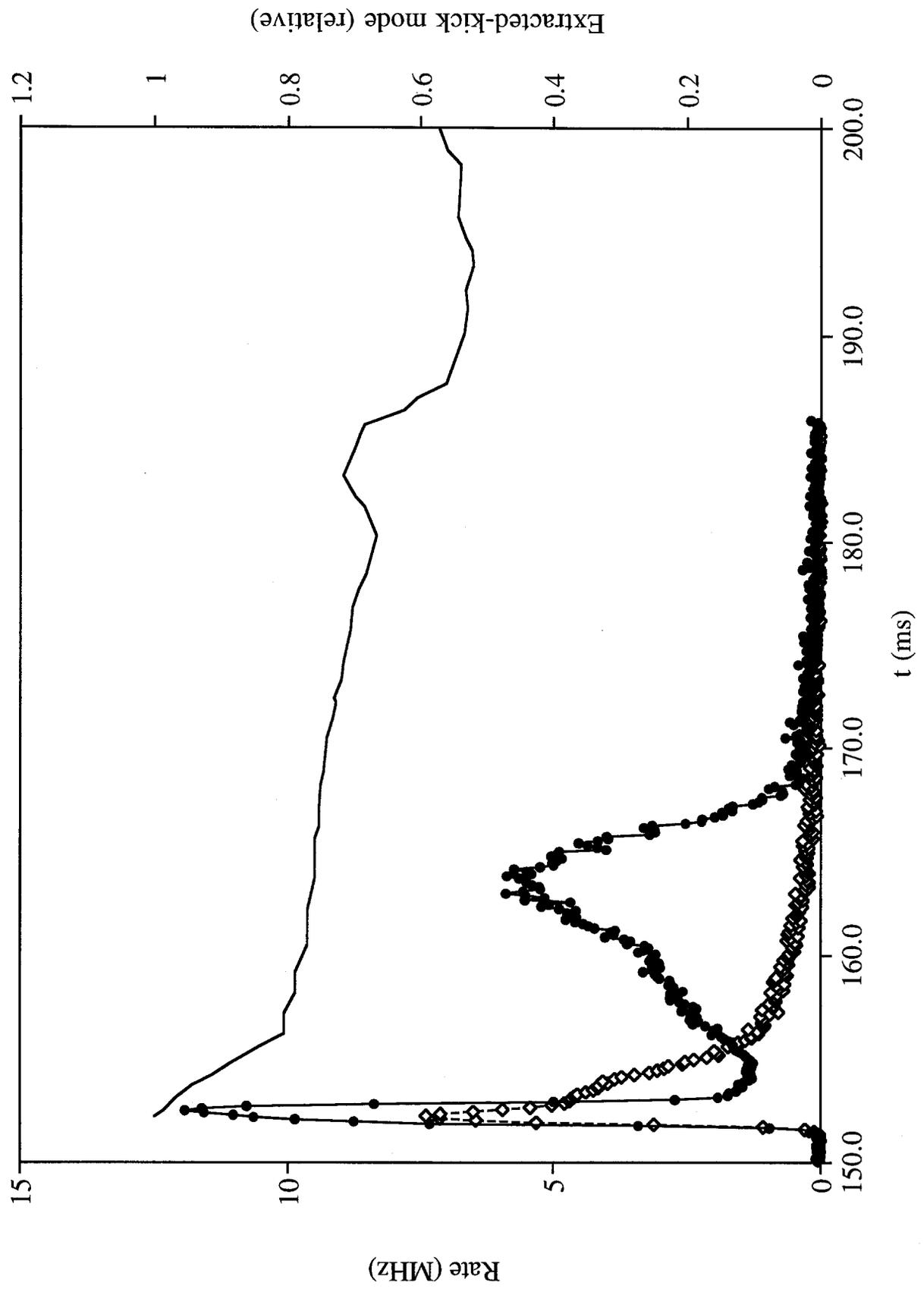

FIG. 23.

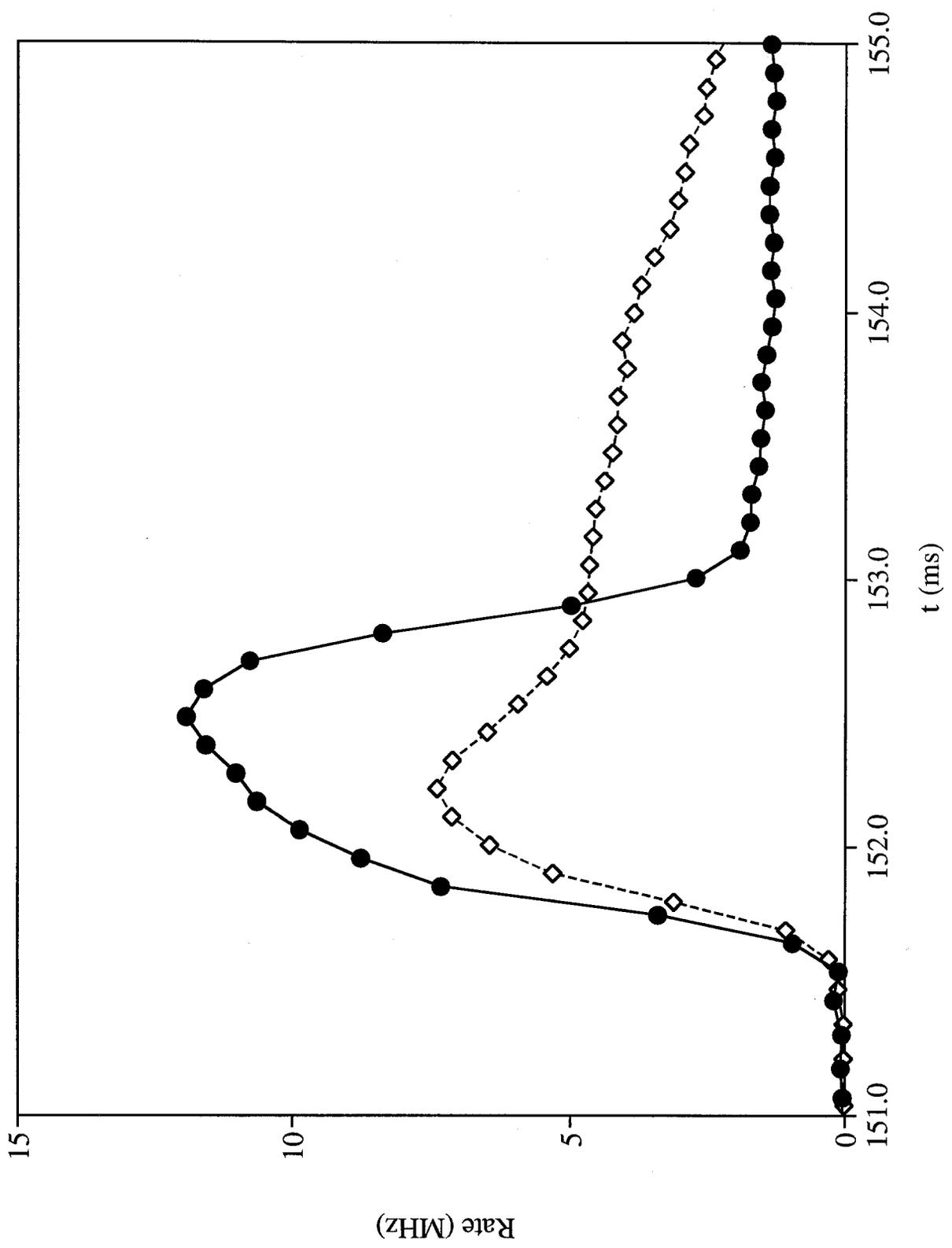

FIG. 24.

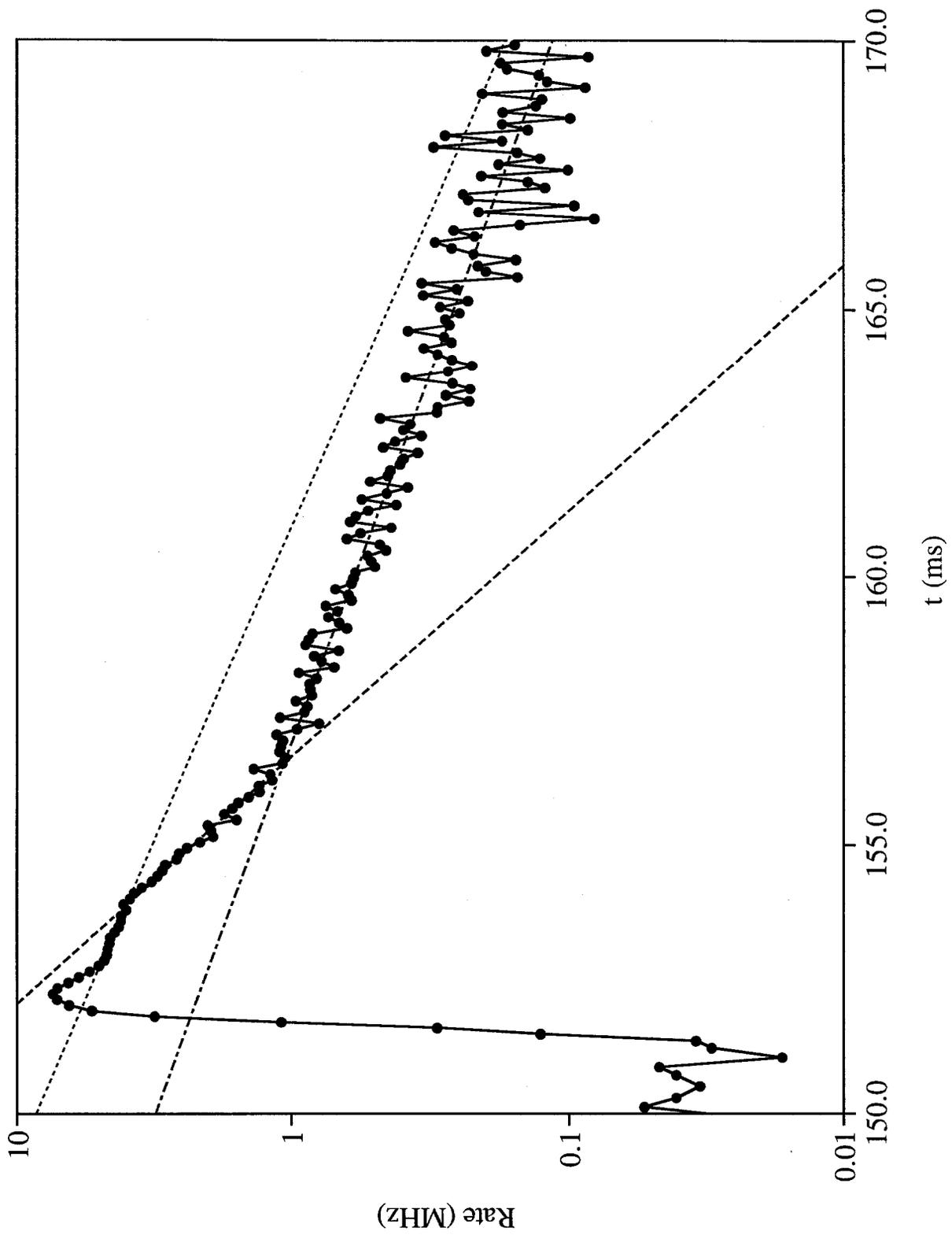

FIG. 25.

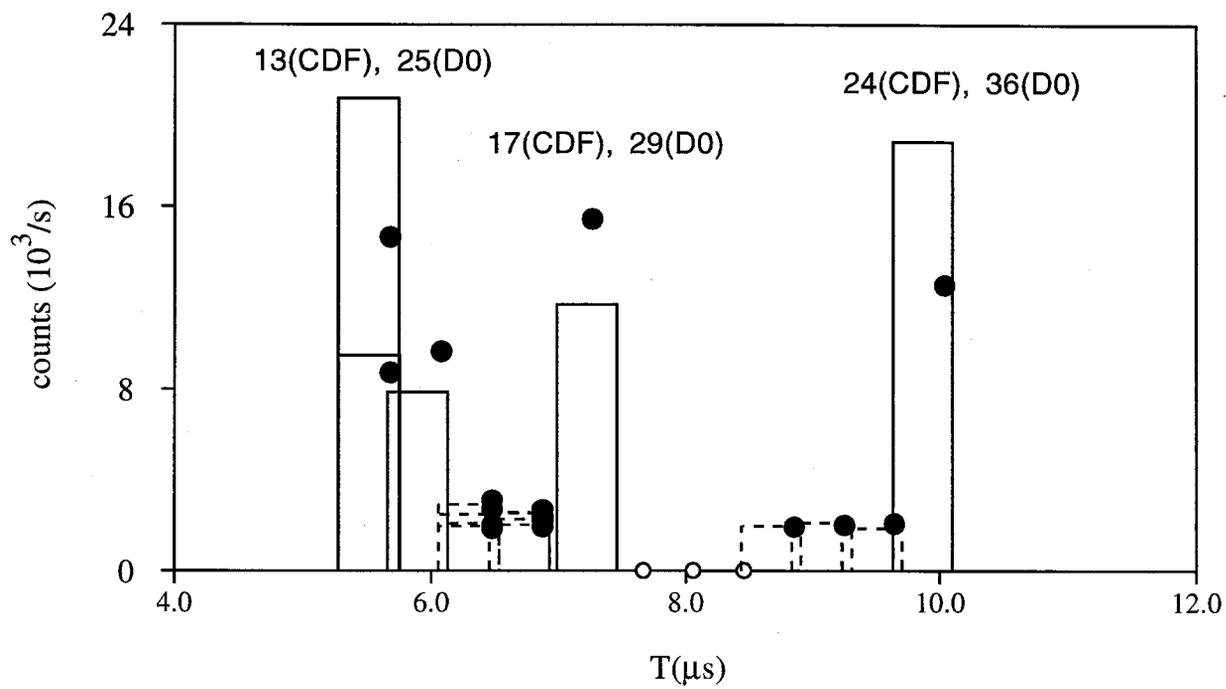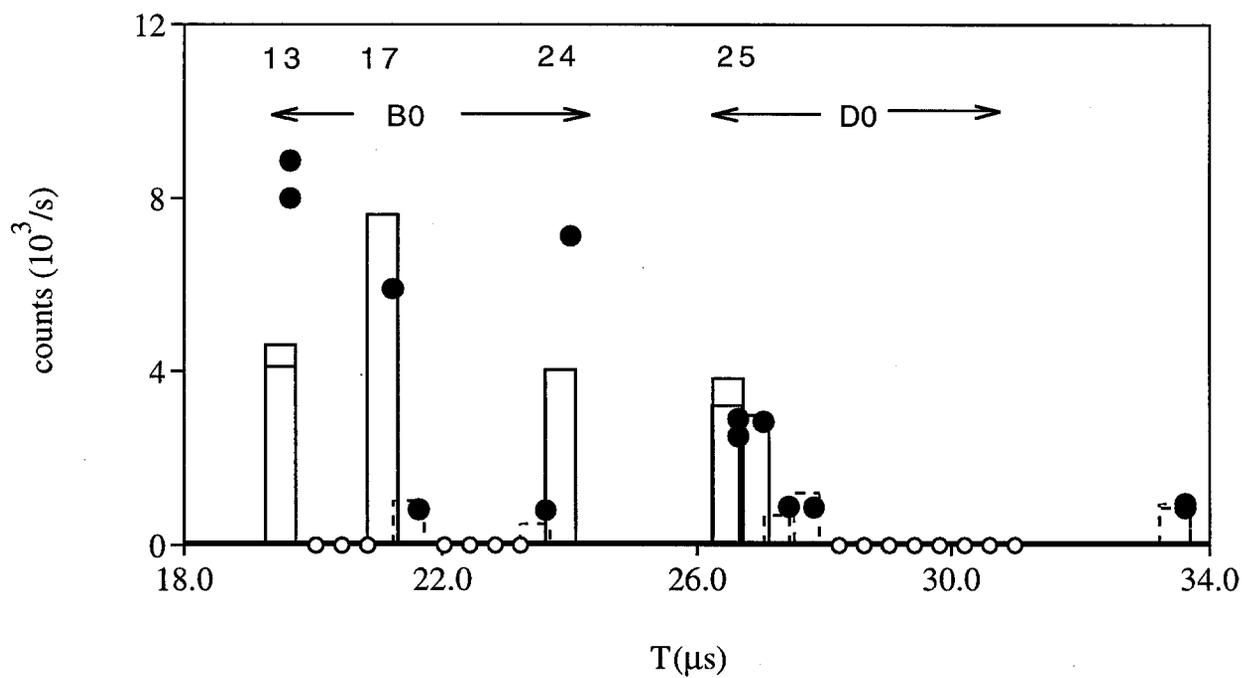

FIG. 26.

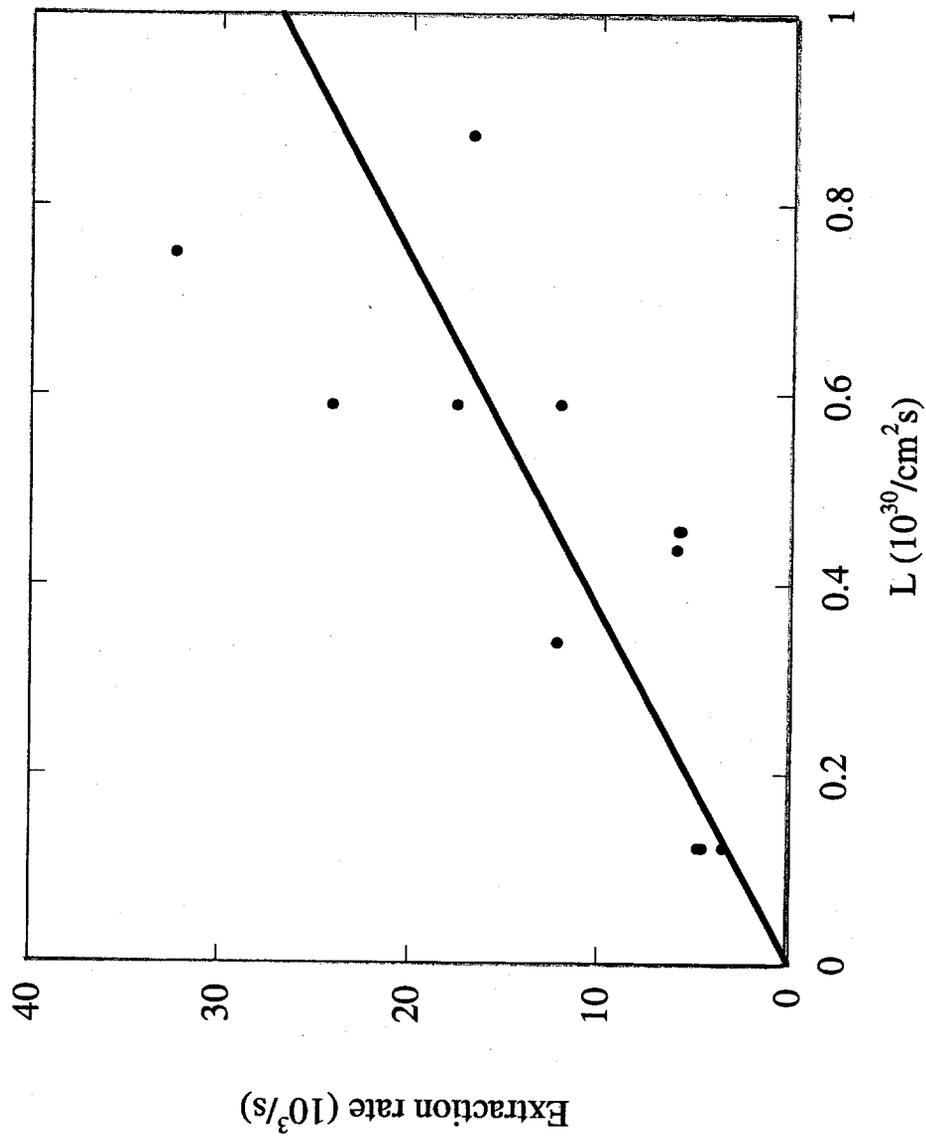

FIG. 27.

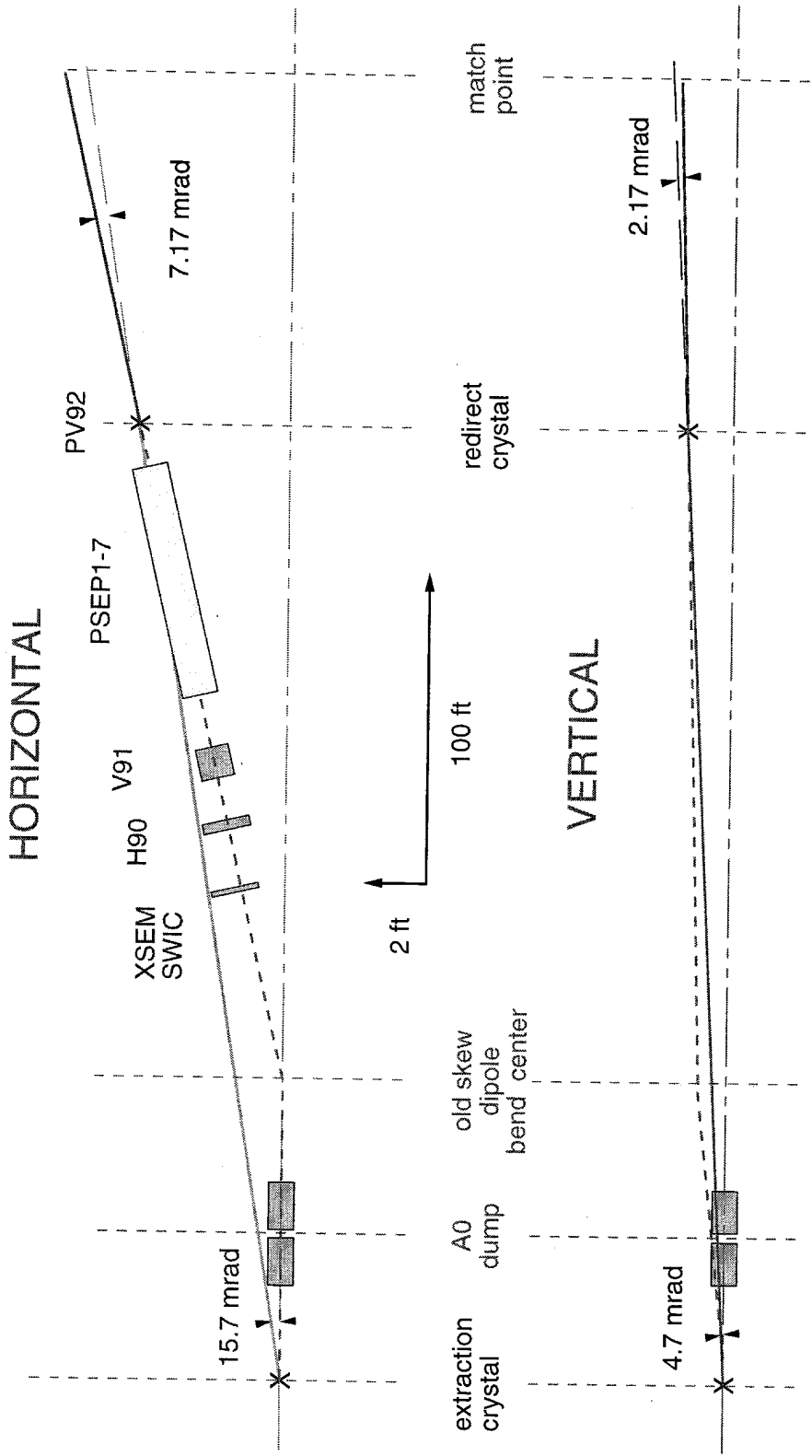

FIG. 28.